\documentclass[preprint,prd,nofootinbib,tightenlines,amsmath,amssymb]{revtex4}
\usepackage[dvipdfmx]{graphicx}
\usepackage{bm}
\usepackage{amsmath}
\usepackage{hhline}
\usepackage{color}
\usepackage{xcolor}
\usepackage{slashed}
\usepackage{relsize}

\newcommand{\Zfive}{{^{}\mathbb{Z}_5\hspace{-0.03cm}}}

\newcommand{\Eq}{&=&}

\begin{document}
\baselineskip=14.5pt \parskip=2.5pt

\vspace*{3em}

\preprint{KUNS-2675}
\preprint{TUM-HEP/1082/17}

\title{A Radiative Neutrino Mass Model with ${}^{}{}^{}$SIMP Dark Matter} 

\author{
Shu-Yu Ho,$^{1,}$\footnote[1]{sho3@caltech.edu}
Takashi Toma,$^{2,}$\footnote[2]{takashi.toma@tum.de}
and 
Koji Tsumura$^{3,}$\footnote[3]{ko2@gauge.scphys.kyoto-u.ac.jp}}
\affiliation{
${}^{1}$Department of Physics, California Institute of Technology, Pasadena, CA 91125, USA \vspace{3pt} \\
${}^{2}$Physik-Department T30d, Technische Universit\"at M\"unchen, James-Franck-Stra\ss{}e, D-85748 Garching, Germany
 \vspace{3pt} \\
${}^{3}$Department of Physics, Kyoto University, Kyoto 606-8502, Japan\vspace{3ex}}


\begin{abstract}
We propose the first viable radiative seesaw model, in which the neutrino masses are induced radiatively 
via the two${}^{}$-loop Feynman diagram involving\,\,Strongly Interacting Massive Particles$\,{}^{}{}^{}$(SIMP). The stability 
of\,\,SIMP dark matter$\,{}^{}$(DM)  is ensured by a ${}^{}\Zfive{}^{}$ discrete symmetry, through which the DM annihilation rate is dominated by the $3 \to 2$ self-annihilating processes. The right amount of thermal relic abundance can 
be obtained with perturbative couplings in the resonant SIMP scenario, while the astrophysical bounds 
inferred from the Bullet cluster and spherical halo shapes can be satisfied. We show that SIMP$\,{}^{}{}^{}$DM is 
able to maintain kinetic equilibrium with thermal plasma until the freeze${}^{}$-out temperature via the Yukawa 
interactions associated with neutrino mass generation. 
\end{abstract}

\maketitle

\section{Introduction}\label{sec:1}
The standard model (SM) of particle physics is an enormously successful theory describing the nature of the universe. Nevertheless, the origin of the non-zero neutrino mass~\cite{Wendell:2010md,Abdurashitov:2009tn, Abe:2012tg, Gando:2010aa} and the identification of dark matter (DM) in the universe~\cite{Begeman:1991iy, Massey:2007wb, Harvey:2015hha, Adam:2015rua} are the lack of explanations in the SM. 


As it is well known, the easiest way to account for tiny neutrino masses
is the canonical seesaw mechanism~\cite{Minkowski:1977sc,
Yanagida:1979as, GellMann:1980vs}, in which heavy right-handed singlet
neutrinos are added to the SM. However, such heavy fermions are very
hard to probe by current colliders. Alternatively, people focus on
radiative seesaw models~\cite{Zee:1980ai, Zee:1985rj, Zee:1985id,
Babu:1988ki}, where neutrino masses are generated at loop level and the
mass scales of the new particles involving in the Feynman diagram can be
lighter than the canonical seesaw mechanism.


On the other hand, a number of well-motivated DM candidates have been
suggested, the most popular among which is the Weakly Interacting
Massive Particles (WIMP) with the mass range spanning from
sub${}^{}{}^{}$-GeV to TeV scale. The WIMP DM is thermally produced
in the early universe, and its relic density is usually determined by
the strength of the ${}^{}{}^{}2{}^{}{}^{}\to{}^{}{}^{}2{}^{}{}^{}$
annihilation cross section of DM into the SM particles. The experimental
investigations for the WIMP DM have null results so far, this motivates physicists to come up with the new perspectives for the DM nature. Recently, a novel idea of DM, Strongly Interacting Massive Particles (SIMP)~\cite{Hochberg:2014dra} has gotten attention and has been explored in the
literature~\cite{Acharya:2017szw,Bernal:2015bla,Bernal:2015lbl,Bernal:2015ova,Bernal:2015xba,Bernal:2017mqb,
Choi:2015bya,Choi:2016tkj,Choi:2016hid,Choi:2017mkk,Cline:2017tka,Dey:2016qgf,Farina:2016llk,Forestell:2016qhc,Halverson:2016nfq,
Hansen:2015yaa,Hochberg:2014kqa,Hochberg:2015vrg,Kamada:2016ois,Kamada:2017tsq,Kuflik:2015isi,Kuflik:2017iqs,
Lee:2015gsa,Lee:2015uva,Pappadopulo:2016pkp,Tsumura:2017knk,Yamanaka:2014pva}. In
comparison with WIMP, the relic abundance of SIMP is determined by the
strength of the ${}^{}{}^{}3{}^{} \to {}^{}2{}^{}{}^{}$ annihilation cross section of DM into itself,
while its mass scale spreads from MeV to sub${}^{}{}^{}$-GeV,
which may be insensitive to present direct searches. 
The annihilation rate for the $3\to2$ process should be larger than the
$2\to2$ annihilation rate to consider SIMP DM instead WIMP. 
In addition, SIMP DM has to be in kinetic equilibrium
with the SM sector until the freeze${}^{}$-out so that the temperature
of the dark sector is the same with that in the SM sector, known as the SIMP condition.
An advantage of SIMP DM opposite to the
WIMP DM is that the SIMP candidate can address some astrophysical issues such as small-scale structure problems~\cite{Elbert:2014bma} and the DM halo separation in Abell 3827 cluster~\cite{Massey:2015dkw, Kahlhoefer:2015vua}.


In the economic point of view, any realistic model beyond the SM should incorporate the above crucial ingredients. The most renowned one possessing these necessary components is Ma's scotogenic model~\cite{Ma:2006km}, in which the WIMP DM is running in the loop diagram to produce the neutrino masses. There are a bunch of studies along this direction~\cite{Ma:2007gq,Krauss:2002px,Aoki:2008av,Gustafsson:2012vj}. 

In this article, we propose a brand-new scheme of the scotogenic model, where the role of WIMP DM is replaced by the SIMP DM. To accomplish our thought, we refer to the resonant SIMP model constructed in Ref.~\cite{Choi:2016tkj} and extend it by introducing more scalars and fermions for neutrino mass generation. Hereafter, we call it $\nu$SIMP model. In this model, the complex scalar is selected as a SIMP DM candidate and is stabilized by a $\Zfive$ symmetry. The resonant effect can reduce the size of the quartic couplings associated with the ${}^{}{}^{}3 {}^{}{}^{}\to {}^{}{}^{}2{}^{}{}^{}$ annihilation processes so that the perturbative bound and the constraints from the Bullet cluster and spherical halo shapes can be satisfied. The SIMP condition can also be fulfilled via the new Yukawa interactions, which connects the dark sector and the SM sector.


The plan of the paper is as follows. In the next section, we introduce the $\nu$SIMP model and give a description of the relevant interactions and masses for the new particles. In Sec.\,\ref{sec:3}, we write down the neutrino mass formula. In Sec.\,\ref{sec:4}, we take into account several experimental and theoretical constraints on the model. In Sec.\,\ref{sec:5}, we evaluate the relic density of the resonant SIMP DM and briefly mention the restrictions from the astrophysical sources. In Sec.\,\ref{sec:6}, we demonstrate the allowed parameter space to make the SIMP condition work. We conclude and summarize our study in Sec.\,\ref{sec:7}. Some lengthy formulas, diagrams, and the benchmark points of the model are put in the appendices.

\section{$\nu$SIMP model 
}\label{sec:2}
To achieve the $\nu$SIMP scenario, we add three vector-like fermions, $N_{1,2,3}{}^{}$, one scalar doublet, 
$\eta{}^{}$, and two complex singlet scalars, $\chi$ and $S$ to the SM, all of which have charges under a conserved 
$\Zfive$ symmetry,\footnote{This discrete symmetry can be realized as a remnant of the U$(1)$ gauge symmetry as discussed in Ref.~\cite{Choi:2016tkj,Ho:2016aye}. A concrete example is given in Appendix A.} while all of the SM particles are $\Zfive$ neutral. The particle contents and the charge 
assignments are summarized in Tab.~\ref{tab:1}.  It follows that the
lightest mass eigenstate (denoted by $X$) 
of the linear combination of ${}^{}\chi$ and the neutral component of ${}^{}\eta$ is stable and can serve as a valid 
SIMP DM candidate.\footnote{In the simplest ${}^{}\mathbb{Z}_3$ SIMP model~\cite{Hochberg:2014dra}, the quartic coupling in the scalar potential is too large to satisfy the bound from perturbativity.}


The renormalizable Lagrangian for the interactions of the scalar particles in this model with one 
another and with the SM gauge bosons is 
\begin{eqnarray}\label{LS}
{\cal L}
\,\,=\,\,
({\cal D}^{\rho}\Phi)^\dagger {}^{} {\cal D}_{\rho}^{} \Phi \,+\,
({\cal D}^{\rho}\eta)^\dagger {}^{} {\cal D}_{\rho}^{} \eta \,+\,
\partial^\rho \chi^\ast {}^{} \partial_\rho {}^{} \chi \,+\, 
\partial^\rho S^\ast {}^{} \partial_\rho {}^{} S \,- {}^{}
{\cal {V}} ~,
\end{eqnarray}
where $\,{\cal D}_\rho\,$ is the SM covariant derivative, and the scalar
potential ${\cal V}$ is
\begin{eqnarray}\label{V}
{\cal V} 
\,\Eq\,
\mu_\Phi^2 {}^{} \Phi^\dagger \Phi \,+\,
\mu_\eta^2 {}^{} \eta^\dagger \eta \,+\,
\mu_\chi^2  {}^{} \chi^\ast \hspace{-0.03cm} \chi \,+\,
\mu_S^2 {}^{} S^\ast \hspace{-0.05cm} S 
\nonumber\\[0.1cm]
&&+\, 
\tfrac{1}{4}\lambda_\Phi (\Phi^\dagger \Phi)^2 +
\tfrac{1}{4}\lambda_\eta (\eta^\dagger \eta)^2 +
\tfrac{1}{4}\lambda_\chi (\chi^\ast \hspace{-0.03cm} \chi)^2 +  
\tfrac{1}{4}\lambda_S (S^\ast \hspace{-0.05cm} S {}^{})^2 
\nonumber\\[0.1cm]
&&+\,
\lambda_{\Phi\eta}   (\Phi^\dagger \Phi)(\eta^\dagger \eta) +
\lambda_{\Phi\eta}'  (\Phi^\dagger \eta)(\eta^\dagger \Phi) +
\lambda_{\Phi\chi}   (\Phi^\dagger \Phi)(\chi^\ast \hspace{-0.03cm} \chi) + 
\lambda_{\Phi S}     (\Phi^\dagger \Phi)(S^\ast \hspace{-0.05cm} S {}^{})
\nonumber\\[0.1cm]
&&+\,
\lambda_{\eta\chi}  (\eta^\dagger \eta)(\chi^\ast \hspace{-0.03cm} \chi) + 
\lambda_{\eta S}    (\eta^\dagger \eta)(S^\ast \hspace{-0.05cm} S {}^{}) +
\lambda_{\chi S}    (\chi^\ast \hspace{-0.03cm} \chi)(S^\ast \hspace{-0.05cm} S {}^{})
\nonumber\\[0.1cm]
&&+\,
\Big[{}^{}{}^{}
\tfrac{1}{2} {}^{} \mu_1\chi^\ast \hspace{-0.05cm} S^2  +
\tfrac{1}{2} {}^{} \mu_2 {}^{} \chi^2 S  +
\tfrac{1}{6}\lambda_3 {}^{} \chi^3 S^\ast +
\tfrac{1}{\sqrt{2}}{}^{} \kappa^{} \upsilon (\Phi^\dag\eta)\chi^\ast +
\text{H.c.}^{}
\Big] ~,
\end{eqnarray}
with ${}^{}{}^{}\upsilon \simeq 246.22\,\,\rm{GeV}{}^{}$ being the vacuum expectation value (VEV) of $\Phi$. The Hermiticity of ${}^{}{}^{}{\cal V}{}^{}$
implies that the parameters in the scalar potential $\mu_{\Phi,\eta,\chi,S}^2{}^{}$,\,$\lambda_{\Phi,\eta,\chi,S,\Phi\eta,\Phi\chi,\Phi S,\eta\chi,\eta S,\chi S}$, and 
$\lambda'_{\Phi\eta}$ must be real. In the later sections, we will
choose $\mu_{1,2}{}^{}, \lambda_3{}^{}$, and $\kappa$ to be real and assume
$\lambda_{\eta, \Phi\eta, \Phi\chi, \Phi S, \eta\chi, \eta S}$ and
$\lambda_{\Phi\eta}'$ are negligible since these quartic couplings are
irrelevant to our numerical analysis.  

\begin{table}[htbp]
\vspace{0.2cm}
\begin{center}
\def\arraystretch{1.4}
\begin{tabular}[c]{|c||c|c||c|c|c|c|}
\hline               
~$\vphantom{|_|^|}$~ & ~$E$~ & ~$\Phi$~ & ~$N_{1,2, 3}$~ & ~$\eta$~ & ~$\chi$~ & ~$S$~ \\[0.06cm]\hline\hline 
~SU(2)$\vphantom{|_|^|}$~       
& ~$\bm{2}$~ & ~$\bm{2}$~ & ~$\bm{1}$~ & ~$\bm{2}$~ & ~$\bm{1}$~ & ~$\bm{1}$~ \\\hline
~U(1)$_{Y}\vphantom{|_|^|}$~      
& ~$-$1/2~ & ~1/2~ & ~0~ & ~1/2~ & ~0~ & ~0~ \\\hline
~$\Zfive$~               
& ~1~ & ~1~ & ~$\omega^2$~ & ~$\omega^2$~ & ~$\omega^2$~ & ~$\omega$~ \\\hline  
\end{tabular}
\caption{Charge assignments of the fermions and scalars in the $\nu$SIMP model, where $E {}^{} = {}^{} \big(\,\nu ~\,\, \ell^-
{}^{}\big){}^{\hspace{-0.05cm}\text{T}}$ is the SM lepton doublet,
$\Phi$ is the SM Higgs doublet, and $\omega\,=\,\exp\big(2\pi i/5\big)$ is the quintic root of unity.}
\vspace{-2cm}
\label{tab:1}
\end{center}
\end{table}

\newpage
After spontaneously symmetry
breaking${}^{}$, the scalar bosons can be parametrized by
\begin{eqnarray}
\Phi 
\,=\, 
\begin{pmatrix}
0 \\[0.1cm] 
\frac{1}{\sqrt2}\big(h+\upsilon\big) 
\end{pmatrix}
~,\quad 
\eta 
\,=\, 
\begin{pmatrix}
\,\eta^+\, \\[0.1cm]  
\,\eta^0\,
\end{pmatrix}
~,
\end{eqnarray}
with ${}^{}h{}^{}$ being the physical Higgs boson. The masses of $\,h,\,S$ and $\,\eta^+$ are then given by
\begin{eqnarray}
m_h^2 \,=\, \tfrac{1}{2}\lambda_\Phi \upsilon^2  ~,\quad
m_S^2 \,=\, \mu_S^2 + \tfrac{1}{2}\lambda_{\Phi S} {}^{} \upsilon^2  ~,\quad
m_{\eta^+}^2 \,=\, \mu_\eta^2 + \tfrac{1}{2}\lambda_{\Phi\eta} \upsilon^2  ~.
\end{eqnarray}
The ${}^{}\kappa \upsilon{}^{}$ term in the scalar potential causes the mixing between the neutral scalars ${}^{}\eta^0$ and $\chi{}^{}$.  In the basis $\scalebox{1.1}{(}\,\eta^0 ~\, \chi\,\scalebox{1.1}{)}{}^{\text{T}}$, the corresponding mass matrix
is written as
\begin{eqnarray}
M^2_{\eta\chi}
\,\equiv\,
\begin{pmatrix}
    m_\eta^2         & m_{\eta\chi}^2 \,     \\[0.15cm]
\,\,m_{\eta\chi}^2 & m_\chi^2 \,
\end{pmatrix}
\,=\,
\begin{pmatrix}
\,\,\mu_\eta^2+\frac{1}{2}\big(\lambda_{\Phi\eta}+\lambda_{\Phi\eta}' \big)\upsilon^2  &  \frac{1}{2}\kappa^{}\upsilon^2 \, 
\\[0.15cm]
\frac{1}{2}\kappa^{}\upsilon^2  &  \mu_\chi^2+\frac{1}{2}\lambda_{\Phi\chi}^{} \upsilon^2\,
\end{pmatrix}~.
\end{eqnarray}
Upon diagonalizing ${}^{}M^2_{\eta\chi}$, we get the mass eigenstates $\,H{}^{}$ and $\,X{}^{}$ and their respective masses ${}^{}m_H{}^{}$ and ${}^{}m_X{}^{}$ given by
\begin{eqnarray}\label{scalarmixing}
\begin{pmatrix}
\,\eta^0\, \\
\,\chi \, 
\end{pmatrix}
\Eq
\begin{pmatrix}
\,c_\xi            &  ~s_\xi~  \\
\,{-}{}^{}s_\xi  &  ~c_\xi~  \\
\end{pmatrix}
\begin{pmatrix}
  H    \\ 
\,X \, 
\end{pmatrix}
\,\equiv\,
{\cal O}_{\eta\chi}
\begin{pmatrix}
  H    \\ 
\,X \, 
\end{pmatrix}
~,\quad
{\cal O}_{\eta\chi}^\text{T}M^2_{\eta\chi}{\cal O}_{\eta\chi}
\,=\,
\text{diag}\big(m_H^2\,,m_X^2\big)  ~,
\nonumber\\[0.2cm]
2{}^{}m_{H,{}^{}X}^2
\Eq
m_{\eta}^2 + 
m_{\chi}^2 \pm 
\sqrt{\big(m_{\eta}^2- m_{\chi}^2{}^{}\big)^{\hspace{-0.05cm}2} +
4{}^{}m_{\eta\chi}^4}
~~,\quad
\sin(2^{}\xi)\,\equiv\,s_{2\xi} 
\,=\,
\frac{\kappa{}^{}\upsilon^2}{m_H^2-m_X^2} ~,
\end{eqnarray}
where $c_\xi = \cos \xi{}^{}, \,s_\xi = \sin \xi{}^{}$, and ${}^{}m_H > m_X{}^{}$. Plugging $\chi = -\,s_\xi H + c_\xi X{}^{}$ into Eq.\eqref{V}, one can extract the relevant interactions for the 3 $\to$ 2 annihilation processes as
\begin{eqnarray}\label{3to2int}
{\cal L}
\,\,\supset\,\,
-\,\tfrac{1}{2} {}^{} \mu_1 c_\xi
\Big[
X^\ast S^2 {}^{} + X \big(S^\ast\big)^{\hspace{-0.05cm}2} \,
\Big]
-
\tfrac{1}{2} {}^{} \mu_2  {}^{} c_\xi^2
\Big[
X^2 S {}^{} + \big(X^\ast\big)^{\hspace{-0.05cm}2} S^\ast {}^{} 
\Big]
-
\tfrac{1}{6}\lambda_3  {}^{} c_\xi^3
\Big[
X^3 S^\ast + \big(X^\ast\big)^{\hspace{-0.05cm}3} S {}^{}
\Big]~.
\end{eqnarray}
These couplings manifest the $\Zfive$ discrete symmetry and can produce the 5${}^{}$-point interactions of $X$ by integrating out the  complex scalar field $S$. To generate the neutrino masses, the additional couplings $\,{\cal L} \,\supset -\, \tfrac{1}{2} {}^{} \mu_2  
\big(s_\xi^2 H^2 - 2 {}^{} c_\xi {}^{} s_\xi X H {}^{} \big)S {}^{}+
\text{H.c.}$ are also required. 
The neutrino masses will be calculated in the next section. 
From Eqs.\eqref{LS}, \eqref{V} and \eqref{scalarmixing}, the Lagrangian describing the invisible decay channels of the $Z$ boson and the Higgs boson is
\begin{eqnarray}\label{invisible}
\,{\cal L} 
\,\,\supset\,\,
\frac{i{}^{}g_\text{w} {}^{} s_\xi^2}{2{}^{}c_\text{w}} 
\scalebox{1.2}{\big(}
X^\ast \partial^\rho X - X \partial^\rho X^\ast
\scalebox{1.2}{\big)}
Z_\rho 
\,-
\scalebox{1.2}{\big(}
\lambda_{\Phi X} {}^{} |X|^2 +
\lambda_{\Phi S} {}^{} |S|^2
\scalebox{1.2}{\big)}
\upsilon {}^{} h ~,\nonumber
\end{eqnarray}
\vspace{-0.8cm}
\begin{eqnarray}
\lambda_{\Phi X} 
{}^{}\equiv\,
\lambda_{\Phi\chi} {}^{}{}^{} c_\xi^2 \, +
\kappa {}^{}{}^{} c_\xi s_\xi + 
\scalebox{1.2}{\big(}
\lambda_{\Phi\eta} \,+ \lambda_{\Phi\eta}' {}^{}
\scalebox{1.2}{\big)} s_\xi^2 ~,
\end{eqnarray}
where ${}^{}g_\text{w}$ is the $\text{SU}(2)$ gauge coupling constant, and $c_\text{w} = \cos\theta_\text{w}$ 
with the weak mixing angle $\theta_\text{w}$. There are also the gauge interactions of the exotic scalars with 
the photon and the weak bosons, which are related to the electroweak precision tests. We collect them 
in Appendix B.


The Lagrangian responsible for the masses and interactions of the vector-like fermions $N_{1,2,3}$ is
\vspace{-0.1cm}
\begin{eqnarray}\label{Yukawa}
{\cal L}_N 
\Eq
- \, M_k {}^{} \overline{N_k} P_L N_k +
{\cal Y}_{jk}
\Big[\,
\overline{\ell^-_j} {}^{}{}^{} \eta^- -\,
\overline{\nu_j} {}^{} \big(c_\xi H^\ast + s_\xi X^\ast\big)
\Big] P_R {}^{} N_k  
\nonumber\\[0.1cm]
&&
-\,\tfrac{1}{2}{\cal Y}^L_{jk} {}^{} \overline{N_j} P_L N_k^\text{c} {}^{} S^\ast -
\tfrac{1}{2}{\cal Y}^R_{jk} {}^{} \overline{N_j} P_R N_k^\text{c} {}^{} S^\ast  \,+\,
\text{H.c.} ~,
\end{eqnarray}
where ${}^{}M_k{}^{}$ represent the Dirac masses, the summation over ${}^{}j, k =1, 2, 3$\, is implicit, the superscript $\text{c}$ refers to the charge conjugation, $P_{R,L} = \frac{1}{2}(1\pm \gamma_5)$, and $\ell_{1,2,3} = e,\mu,\tau$. Explicitly, the   
Yukawa couplings ${}^{}{\cal Y}_{rk}$ and  ${}^{}{\cal Y}^{L,R}_{rk}$ are of the forms as
\begin{eqnarray} \label{yukawa}
{\cal Y} 
\,=\,
\left(\begin{array}{ccc} 
Y_{e 1}     & Y_{e 2}      & Y_{e 3}     \\
Y_{\mu 1} & Y_{\mu 2}  & Y_{\mu 3}  \\ 
Y_{\tau 1} & Y_{\tau 2}  & Y_{\tau 3}
\end{array}\right) ~, \quad
{\cal Y}^{L,R} 
\,=\,
\left(\begin{array}{ccc} 
{\cal Y}^{L,R} _{11}    & {\cal Y}^{L,R} _{12}  & {\cal Y}^{L,R} _{13} \\[0.15cm]
{\cal Y}^{L,R} _{21}    & {\cal Y}^{L,R} _{22}  & {\cal Y}^{L,R} _{23} \\ [0.15cm]
{\cal Y}^{L,R} _{31}    & {\cal Y}^{L,R} _{32}  & {\cal Y}^{L,R} _{33}
\end{array}\right) ~, 
\end{eqnarray}
where $\,Y_{\ell_j k} = {\cal Y}_{jk}{}^{}$. 


\section{Radiative Neutrino mass 
}\label{sec:3}
\begin{figure}[t]
\vspace{-0.2cm}
\begin{center}
\includegraphics[scale=0.45]{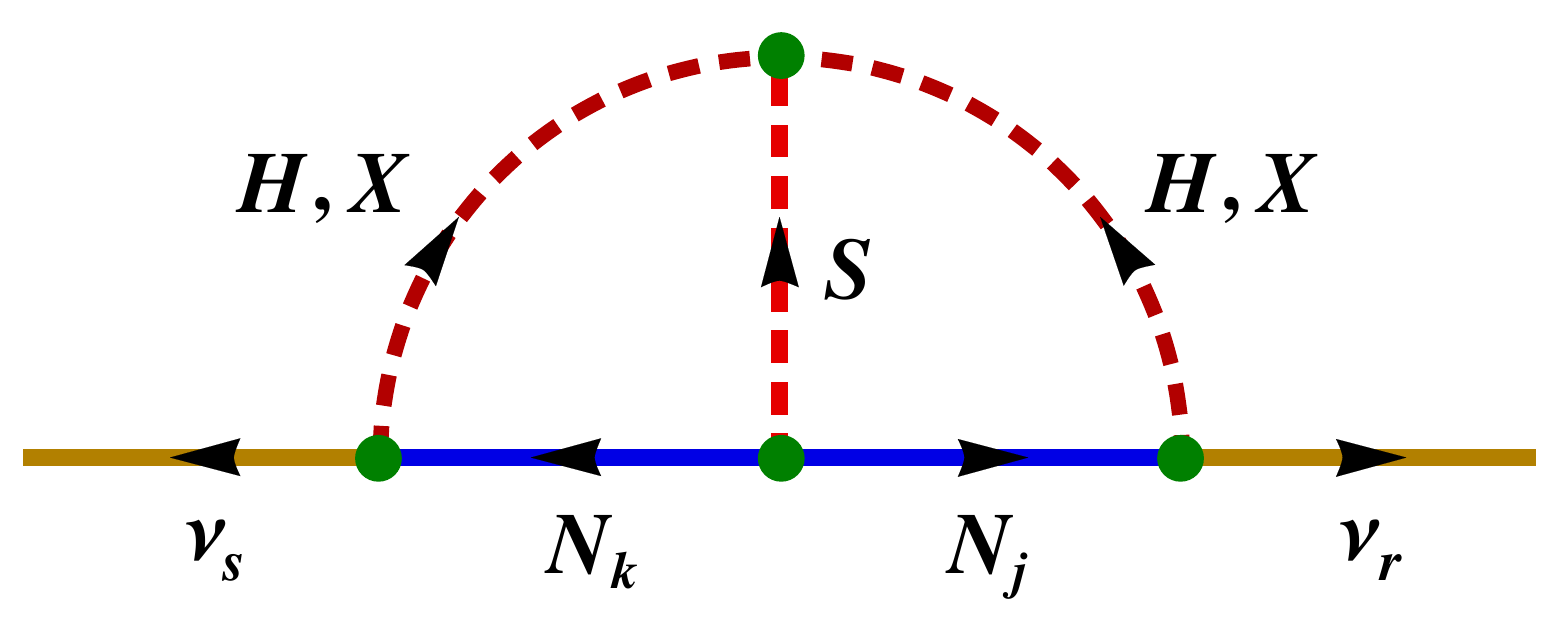}
\vspace{-0.5cm}
\caption{Feynman diagrams for neutrino mass generation at the two${}^{}$-loop level.}
\label{fig:two_loop}
\end{center}
\vspace{-0.5cm}
\end{figure}

In the $\nu$SIMP model, the neutrinos acquire mass radiatively through two${}^{}$-loop diagrams with internal 
$H, X, S$, and $N_k$ as shown in Fig.~\ref{fig:two_loop}. 
The resulting neutrino mass matrix defined by ${\cal
L}_{\nu}  = -\frac{1}{2}({\cal M}_\nu)_{rs}{}^{}\overline{\nu_r}{}^{}\nu_s^\text{c} + \text{H.c.}$ is given as~\cite{Aoki:2014cja, Ding:2016wbd}
\begin{eqnarray}
\big({\cal M}_\nu\big)_{rs} 
\,=\, 
\frac{\mu_2 {}^{}{}^{} {\cal Y}_{rj} {}^{} {\cal Y}_{sk} {}^{} s_{2\xi}^2}{4(4\pi)^4}
\Big({\cal Y}_{jk}^L {}^{} {\cal C}_{jk}^L + {\cal Y}_{jk}^R {}^{} {\cal C}_{jk}^R \Big) ~,
\label{eq:nu-mass}
\end{eqnarray}
where the loop functions are 
\begin{eqnarray}
{\cal C}_{jk}^L 
\Eq
\mathop{\mathlarger{\int}_0^1}\hspace{-0.1cm}d\hat{u} {}^{} d\hat{v} {}^{} d\hat{w}\,\frac{\delta\big(\hat{u}+\hat{v}+\hat{w}-1\big)}{1-\hat{w}}
\Bigg[\,
{\cal I}_L\bigg(\frac{m_X^2}{M_k^2},\frac{m_{X j S}^2}{M_k^2} \bigg) \hspace{-0.05cm}-{}^{}
{\cal I}_L\bigg(\frac{m_X^2}{M_k^2},\frac{m_{H j S}^2}{M_k^2} \bigg) 
\nonumber\\[0.1cm]
&&\hspace{5.6cm}-\,
{\cal I}_L\bigg(\frac{m_H^2}{M_k^2},\frac{m_{X j S}^2}{M_k^2} \bigg) \hspace{-0.05cm}+{}^{}
{\cal I}_L\bigg(\frac{m_H^2}{M_k^2},\frac{m_{H j S}^2}{M_k^2} \bigg) 
\Bigg]
~,
\nonumber\\[0.1cm]
{\cal C}_{jk}^R 
\Eq
\frac{M_j}{M_k} 
\mathop{\mathlarger{\int}_0^1}\hspace{-0.1cm}d\hat{u} {}^{} d\hat{v} {}^{} d\hat{w}\,\frac{\delta\big(\hat{u}+\hat{v}+\hat{w}-1\big)}{\hat{w}(1-\hat{w})}
\Bigg[\,
{\cal I}_R\bigg(\frac{m_X^2}{M_k^2},\frac{m_{X j S}^2}{M_k^2} \bigg) \hspace{-0.05cm}-{}^{}
{\cal I}_R\bigg(\frac{m_X^2}{M_k^2},\frac{m_{H j S}^2}{M_k^2} \bigg) 
\nonumber\\[0.1cm]
&&\hspace{6.35cm}-\,
{\cal I}_R\bigg(\frac{m_H^2}{M_k^2},\frac{m_{X j S}^2}{M_k^2} \bigg) \hspace{-0.05cm}+{}^{}
{\cal I}_R\bigg(\frac{m_H^2}{M_k^2},\frac{m_{H j S}^2}{M_k^2} \bigg) 
\Bigg] ~,
\end{eqnarray}
with
\begin{eqnarray}
{\cal I}_L(a,b) \,=\, \frac{a^2 \ln a}{(1-a)(a-b)} + \frac{b^2 \ln b}{(1-b)(b-a)} ~,\quad
{\cal I}_R(a,b) \,=\, \frac{a \ln a}{(1-a)(a-b)} + \frac{b \ln b}{(1-b)(b-a)}  ~,\quad\nonumber
\end{eqnarray}
\vspace{-0.4cm}
\begin{eqnarray}
m_{X j S}^2 \,=\, \frac{\hat{u}{}^{}{}^{}m_X^2+\hat{v}{}^{}M_j^2+\hat{w}{}^{}{}^{}m_S^2}{\hat{w}(1-\hat{w})} ~,\quad
m_{H j S}^2 \,=\, \frac{\hat{u}{}^{}{}^{}m_H^2+\hat{v}{}^{}M_j^2+\hat{w}{}^{}{}^{}m_S^2}{\hat{w}(1-\hat{w})} ~.
\end{eqnarray}
The mass matrix in Eq.\,(\ref{eq:nu-mass}) is diagonalized by the
Pontecorvo${}^{}{}^{}$-Maki${}^{}$-Nakagawa${}^{}$-Sakata (PMNS) matrix ${}^{}\text{U}_\mathrm{PMNS}{}^{}$ as ${}^{}\text{U}_\mathrm{PMNS}^\dagger {}^{}{}^{} \mathcal{M}_{\nu}  \text{U}_\mathrm{PMNS}^\ast=\mathrm{diag}\big(m_{\nu_1},m_{\nu_2},m_{\nu_3}\big)$.
The mixing angles in the PMNS matrix and neutrino mass eigenvalues are given
by the global fitting to the neutrino oscillation data~\cite{Gonzalez-Garcia:2014bfa}. 

\newpage
As we will discuss in Sec.\,\ref{sec:6}, the order of magnitude of the
Yukawa couplings is ${}^{}{\cal Y}_{jk} \sim {\cal O}(0.01-1)$ with 
$0.1 \,\text{GeV} \lesssim M_k \lesssim 1 \,\text{GeV}$ when the SIMP
condition is imposed. In the next section, we will also show that the
size of the mixing angle should be ${}^{}s_\xi {}^{} \lesssim {}^{} 0.06{}^{}$ due
to the constraints from the invisible decays of  the $Z$ boson and the
Higgs boson. Moreover, in order to satisfy perturbative bounds on the quartic couplings and 
the observed relic density of DM, we find that the cubic coupling ${}^{}\mu_2 \sim {\cal O}(100 \,\,\text{MeV})$. Accordingly, if one takes ${\cal Y}_{jk} \sim 0.1, \,s_\xi \sim 0.05 , \,\mu_2 \sim 100 \,\text{MeV}, \,{\cal Y}^{L,R}_{jk} \sim 0.1$, and 
${\cal C}{}^{L,R}_{jk} \sim 1$, the correct neutrino mass scale ${}^{}m_\nu \sim 0.1 \,\text{eV}$ can be arrived. To make this model more reliable, we display the benchmark points in Appendix C.


\section{Constraints}\label{sec:4}
\begin{figure}[t]
\begin{center}
\includegraphics[scale=0.45]{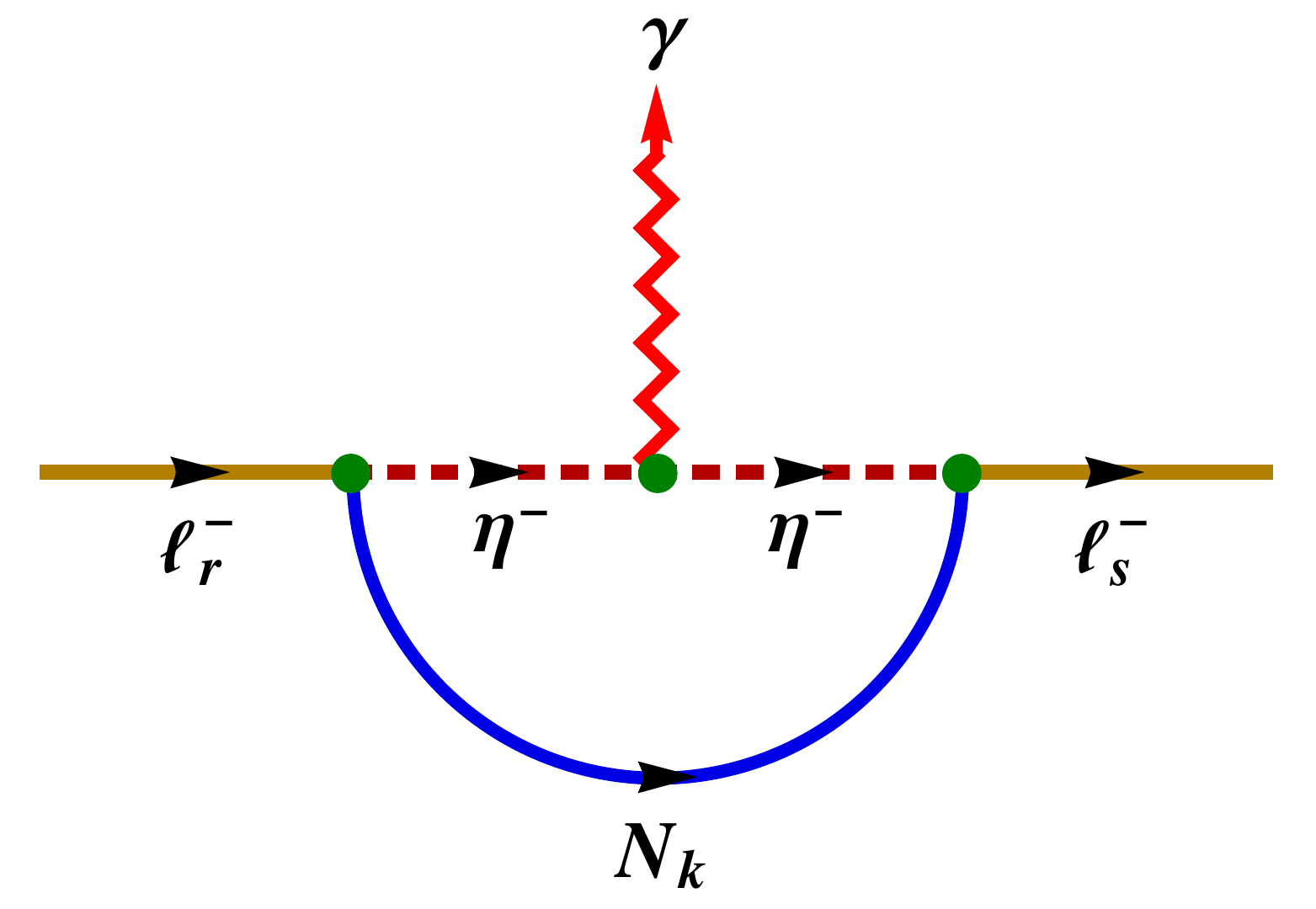}
\vspace{-0.4cm}
\caption{Feynman diagram of flavor-changing radiative decay ${}^{}\ell_r \to \ell_s \gamma{}^{}$.}
\label{fig:LFV}
\end{center}
\vspace{-0.5cm}
\end{figure}

There are various experimental and theoretical restrictions on the masses and couplings of the new particles in the 
$\nu$SIMP scenario. Experimentally, the flavor-changing radiative decay ${}^{}\ell_r \to \ell_s \gamma{}^{}$ constrains 
the Yukawa couplings ${\cal Y}_{rk}{}^{}$. The Feynman diagram depicted such decay process
is shown in Fig.~\ref{fig:LFV}. The branching fraction of the decay process is
\begin{eqnarray}
{\cal B}\big(\ell_r\rightarrow \ell_s\gamma\big)
\,=\,
\frac{3{}^{}\alpha \, {\cal B}\big(\ell_r\rightarrow \ell_s\nu_r \nu_s \big)}{64{}^{}\pi{}^{} G_\text{F}^2{}^{}m_{\eta^+}^4}\,
\Bigg|\sum_{k\,=\,1}^{3}
{\cal Y}_{rk}^\ast{\cal Y}_{sk}\,
{\cal F}\bigg(\frac{M_k^2}{m_{\eta^+}^2}\bigg)
\Bigg|^2
~,
\end{eqnarray}
where the fine structure constant $\alpha$, the Fermi constant
$G_\text{F}$ and the loop function ${\cal F}(z)$ are 
\begin{eqnarray}
\alpha \,=\, \frac{\hat{e}^2}{4\pi} ~,\quad
G_\text{F} \,=\, \frac{1}{\sqrt{2}{}^{}{}^{}\upsilon^2} ~,\quad
{\cal F}(z) \,=\, \frac{1-6{}^{}z+3{}^{}z^2+2{}^{}z^3-6{}^{}z^2\ln z}{6{}^{}(1-z)^4} ~,
\end{eqnarray}
with $\hat{e}$ the electromagnetic charge. The most stringent experimental limit on the ${}^{}\mu \to e \gamma{}^{}$ process comes from the MEG Collaboration~\cite{TheMEG:2016wtm}. The up-to-date upper bound on its branching ratio is $\,{\cal B}(\mu \to e \gamma) < 4.2 {}^{}\times 10^{-13}$. If we take ${}^{}m_{\eta^+} \hspace{-0.06cm} \sim 300 \, \text{GeV}$ and ${\cal F}(M_k^2/m_{\eta^+}^2) =1/6$ with $m_{\eta^+} \hspace{-0.06cm} \gg M_k{}^{}$,\footnote{If $\eta^\pm$ decays dominantly into electron or muon, $m_{\eta^+} \hspace{-0.06cm} {}^{} \gtrsim {}^{}270 ~ \text{GeV}$ is required in order to avoid the constraint from the left-handed slepton search~\cite{Aad:2014vma}.} the Yukawa couplings are limited in ${\cal Y}_{rk} \lesssim 0.02$ 
which is in conflict with the range mentioned in the previous
section. The simplest solution to evade this severe constraint is to
assume a diagonal Yukawa matrix ${}^{}{\cal Y}$. In this solution, 
the pattern of neutrino mixing is pinned down by the structures of the other Yukawa matrices ${\cal Y}^{L,R}$ and the mass hierarchy of the vector-like fermions.\footnote{If $m_\tau > m_{e,\mu}+ 2 M_k$, the new decay modes $\tau \to (e,\mu) \bar{N}N' \to (e,\mu) {}^{}  \nu\nu'\bar{X}X$ open and would contribute to $\tau \to (e,\mu) + \text{missing energy}$. However, this constraint is not so stringent.}

At one${}^{}{}^{}$-loop level, the presence of $\eta^{\pm}$ and $N_k$ also induces a contribution to the anomalous
magnetic moment $a^{}_{\ell_j}\hspace{-0.05cm}$ of charged lepton $\ell_j$ given by
\begin{eqnarray}\label{muon}
\Delta {}^{} a^{}_{\ell_j} 
=\,
-{}^{}{}^{}
\frac{m_{\ell_j}^2}{16{}^{}\pi^2m_{\eta^+}^2}
\sum_{k\,=\,1}^{3} 
\big| {\cal Y}_{jk} \big|^2
{\cal F}\bigg(\frac{M_k^2}{m_{\eta^+}^2}
\bigg) ~.
\end{eqnarray}
In particular, the current experimental value for the muon anomalous
magnetic moment has more than $3{}^{}\sigma$ deviation from the SM prediction$\,\,$: \hspace{-0.1cm}$a^\text{exp}_\mu - a^\text{SM}_\mu = (288 \pm
80)\times 10^{-11}$\,\cite{Agashe:2014kda}. 
Since the new contribution given by Eq.\eqref{muon} is negative, we then require $|\Delta {}^{} a^{}_\mu| < 8 \times 10^{-10}{}^{}$, this gives an upper bound for the Yukawa coupling ${\cal Y}_{rk}{}^{}$. 
For example, by taking $m_{\eta^+} \hspace{-0.05cm}  \sim 300 \,\text{GeV}$ and $m_{\eta^+} \hspace{-0.05cm} \gg M_k{}^{}$, the Yukawa couplings are limited in ${}^{}{}^{}{\cal Y}_{rk} \lesssim {\cal O}(1)$ which is less stringent compared to the constraints from the flavor-changing radiative decay.
%

In our study, we suggest that the lightest complex scalar $X$ is the SIMP DM candidate. Since the mass scale of \,SIMP DM 
is MeV to sub${}^{}{}^{}$-GeV, there is then a new physics contribution to the invisible decay width of the $Z$ boson and the Higgs boson, namely ${}^{}\Gamma_{Z,\,h \,\to\, \text{inv}}^\text{new} = \Gamma\scalebox{1.1}{(}Z,h \to X\bar{X}\scalebox{1.1}{)}$. The present experimental bounds on these invisible decay widths are ${}^{}\Gamma_{Z \,\to\, \text{inv}}^\text{new} < 2
~\text{MeV}$ (at the 95\%\,C.L.)~\cite{Carena:2003aj} and
${}^{}\Gamma_{h \,\to\, \text{inv}}^\text{new} \lesssim
0.78~\text{MeV}$. To derive the latter one, we interpret ${\cal
B}^\text{new}_{h \,\to\,\text{inv}} = {\cal
B}^{{}^{}\text{exp}}_\text{BSM} < 0.16$~\cite{Khachatryan:2016vau}
reported by the ATLAS and CMS combined measurements and adopt the SM
Higgs width $\Gamma_h^\text{SM} = 4.08
~\text{MeV}$~\cite{Heinemeyer:2013tqa} at $m_h =
125.1\,\text{GeV}$\,\,\cite{Aad:2015zhl}. From Eqs.\eqref{scalarmixing} and \eqref{invisible},
these upper limits consequently are translated into $|s_\xi| \hspace{-0.015cm} \lesssim \hspace{-0.015cm}  0.4$ and $|s_\xi| \hspace{-0.015cm} \lesssim \hspace{-0.015cm} 0.165 \, (m_H/100\,{}^{}\text{GeV})^{-1}$, respectively. It turns out that the constraint from the Higgs
invisible decay width is much stronger than the $Z$ boson one. For instance, by choosing $m_H \sim 300 \,{}^{} \text{GeV}$, we then reach the upper bound $|s_\xi| \lesssim 0.06$.


The scalar masses are constrained by the oblique parameters due to their
modifications to the SM gauge boson
propagators~\cite{Peskin:1991sw}. From Eq.\eqref{gaugeint} and
Fig.~\ref{fig:EW} in Appendix~\ref{sec:appa}, those parameters are calculated as
\begin{eqnarray}
\Delta {\cal S} 
\Eq 
\frac{1}{12{}^{}\pi}
\scalebox{1.2}{\Big[}
c_\xi^2 \ln \hspace{-0.08cm} \scalebox{1.2}{\big(} m_H^2/m_{\eta^+}^2 \scalebox{1.2}{\big)} \hspace{-0.05cm}+
s_\xi^2 \ln \hspace{-0.08cm} \scalebox{1.2}{\big(} m_X^2/m_{\eta^+}^2 \scalebox{1.2}{\big)} \hspace{-0.05cm}+
c_\xi^2 {}^{} s_\xi^2 {}^{} {}^{} G\scalebox{1.2}{\big(}m_X^2,m_H^2\scalebox{1.2}{\big)}
\scalebox{1.2}{\Big]}~,
\nonumber\\[0.15cm]
\Delta {\cal T}
\Eq 
\frac{1}{8{}^{}\alpha{}^{}\pi^2{}^{}\upsilon^2}
\scalebox{1.2}{\Big[}
c_\xi^2 {}^{} F \scalebox{1.2}{\big(}m_{\eta^+}^2,m_H^2\scalebox{1.2}{\big)} \hspace{-0.05cm} +
s_\xi^2 {}^{} F \scalebox{1.2}{\big(}m_{\eta^+}^2,m_X^2\scalebox{1.2}{\big)} \hspace{-0.05cm}-
c_\xi^2 s_\xi^2 {}^{} F\scalebox{1.2}{\big(} m_X^2,m_H^2\scalebox{1.2}{\big)} 
\scalebox{1.2}{\Big]} ~,
\nonumber\\[0.2cm]
\Delta {}^{} {\cal U}
\Eq
\frac{1}{12{}^{}\pi}
\scalebox{1.2}{\Big[}
c_\xi^2 {}^{} G\scalebox{1.2}{\big(}m_{\eta^+}^2,m_H^2\scalebox{1.2}{\big)} \hspace{-0.05cm} +
s_\xi^2 {}^{} G\scalebox{1.2}{\big(}m_{\eta^+}^2,m_X^2\scalebox{1.2}{\big)} \hspace{-0.05cm} -
c_\xi^2 {}^{} s_\xi^2 {}^{}{}^{} G\scalebox{1.2}{\big(}m_X^2,m_H^2\scalebox{1.2}{\big)}
\scalebox{1.2}{\Big]} ~,
\end{eqnarray}
where the loop functions are given by
\begin{eqnarray}
F(a,b) \Eq \frac{a+b}{2} - \frac{a {}^{} b}{a-b} \ln \hspace{-0.08cm} \bigg(\frac{a}{b}\bigg) ~,\nonumber
\end{eqnarray}
\vspace{-0.5cm}
\begin{eqnarray}
G(a,b) \Eq 
\frac{22{}^{}a{}^{} b - 5{}^{}a^2 - 5{}^{}b^2}{3(a-b)^2} 
+
\frac{(a+b)\scalebox{1.2}{(}a^2-4a {}^{} b +b^2\scalebox{1.2}{)}}{(a-b)^3} 
\ln \hspace{-0.08cm} \bigg(\frac{a}{b}\bigg) ~.
\end{eqnarray}
Since we are interested in the scale that the mass ${}^{}m_X$ is below electroweak scale, one may think that more general definitions of the oblique parameters may be used~\cite{Burgess:1993mg}. However, we have checked the difference is not important because the most stringent constraint comes from the $ {\cal T}$-parameter whose definition does not change even for 
below electroweak scale. The current constraints are given in Ref.~\cite{Baak:2012kk,Baak:2014ora} as $\Delta {\cal S} = 0.05 {}^{} \pm {}^{} 0.11$, $\Delta {\cal T} = 0.09 {}^{} \pm {}^{} 0.13$, $\Delta {}^{} {\cal U} = 0.01 {}^{} \pm {}^{} 0.11$ with the correlation coefficients $0.90\,$(between $\Delta {\cal S}$ and $\Delta {\cal T}\,$), $-\,0.59\,$(between $\Delta {\cal S}$ and $\Delta {}^{} {\cal U}{}^{}{}^{}$), and $\hspace{-0.05cm}-\,0.83\,$(between $\Delta {\cal T}$ and $\Delta {}^{} {\cal U}{}^{}{}^{}$). These limits imply that the heavier neutral component $H$ and the charged component $\eta^+$ should be nearly degenerate $\big( m_H \approx m_{\eta^+}\hspace{-0.03cm}\big)$ in the case of $s_\xi \ll 1$ and $m_X \ll m_H,{}^{}m_{\eta^+} {}^{}.
$


Theoretically, the quartic parameters $\lambda_j\hspace{-0.03cm}$ are subject to the conditions of vacuum stability and perturbativity.
To ensure the vacuum to be stabilized at large field values, we demand~\cite{Choi:2016tkj}
\begin{eqnarray}
\lambda_{X,{}^{}S} > 0 ~,~~
\lambda_{XS} > -^{}\tfrac{1}{2} \sqrt{\lambda_X \lambda_S} ~,~~
|\lambda_3| 
< \, 
\sqrt{\frac{18 {}^{} \lambda_X \lambda_S \lambda_{XS} - 8 {}^{} \lambda_{XS}^3
+
\scalebox{1.1}{\big(}
4{}^{}\lambda_{XS}^2 + 3{}^{}\lambda_X \lambda_S
\scalebox{1.1}{\big)}^{\hspace{-0.06cm}3/2}
}{3{}^{}\lambda_X}}
~,
\end{eqnarray}
where $\lambda_X \equiv \lambda_\chi {}^{}{}^{} c_\xi^4 \,\approx \lambda_\chi {}^{}$, and 
${}^{}\lambda_{XS} \equiv \lambda_{\chi S} {}^{} c_\xi^2 \,\approx \lambda_{\chi S}{}^{}$ because of the smallness of the
mixing angle $\xi{}^{}$. 
In the previous work~\cite{Choi:2016tkj}, a condition of perturbativity
on the quartic couplings has been taken, which corresponds to
${}^{}\lambda_{X,{}^{}S} < 16{}^{}\pi{}^{}$ in our convention in Eq.\eqref{V}.
However, this upper bound seems to be overly optimistic when the RG
running is considered. Instead, we force the relatively conserved
conditions $\lambda_{X,{}^{}S} < 4{}^{}\pi$ in our numerical
work. Furthermore, since $X$ plays the role of DM, it should not develop the VEV.
The sufficient conditions to guarantee $\langle X \rangle = {}^{} 0$ (as well as $\langle S \rangle = 0$)
are given by
\begin{eqnarray}\label{novev}
\lambda_X > \frac{\mu_2^2}{m_S^2} ~,\quad
\lambda_S > \frac{\mu_1^2}{m_X^2}  ~,\quad
\lambda_{X S} > 0 ~,
\end{eqnarray}
here we have assumed ${}^{}\lambda_3 = 0{}^{}$ for simplicity.\footnote{One can find the necessary conditions to ensure $\langle X\rangle = 0$ by using the method in the literature~\cite{Kannike:2012pe}. However, the analytical result is too long to read.}

\section{Resonant SIMP DM and relic abundance}\label{sec:5}
In order to estimate the thermal relic abundance of SIMP DM,
we have to solve the Boltzmann equation of the DM number density $n_\text{DM} = n_X + n_{\bar{X}} = 2{}^{}n_X$\,(we assume there is no asymmetry between particles $X$ and $\bar{X}{}^{}$) as follows 
\begin{eqnarray}
\frac{d^{}n_\text{DM}}{d{}^{}t} + 3 {}^{}{}^{} {\cal H} {}^{} n_\text{DM}
\Eq
-\,\big\langle \sigma_{3 \to 2}\upsilon_\text{rel}^2 {}^{} \big\rangle
\scalebox{1.2}{\big(}
n_\text{DM}^3-n_\text{DM}^2 n_\text{DM}^\text{eq}
\scalebox{1.2}{\big)}
~,
\end{eqnarray}
with ${}^{} {\cal H}{}^{}$ being the Hubble parameter, $n_\text{DM}^\text{eq}$ the DM number density at the chemical equilibrium, and $\langle \sigma_{3 \to 2}\upsilon_\text{rel}^2 \rangle \equiv \frac{1}{24}\langle \sigma_{XXX \to \bar{X}\bar{X}}{}^{}\upsilon_\text{rel}^2 \rangle$ the thermal averaged effective $3 \to 2$ annihilation cross section.\footnote{The definition of the effective $3 \to 2$ annihilation cross section depends on the model. For example, in the $\mathbb{Z}_3$ SIMP model~\cite{Choi:2015bya}, $\big\langle \sigma_{3 \to 2}\upsilon_\text{rel}^2 {}^{} \big\rangle \equiv \frac{1}{24} \big\langle \sigma_{XXX \to X\bar{X}}{}^{}\upsilon_\text{rel}^2 {}^{} \big\rangle+\frac{1}{8}\big\langle \sigma_{XX\bar{X} \to \bar{X}\bar{X}}{}^{}\upsilon_\text{rel}^2 {}^{} \big\rangle$.} 
\begin{figure}[t]
\begin{center}
\includegraphics[scale=0.25]{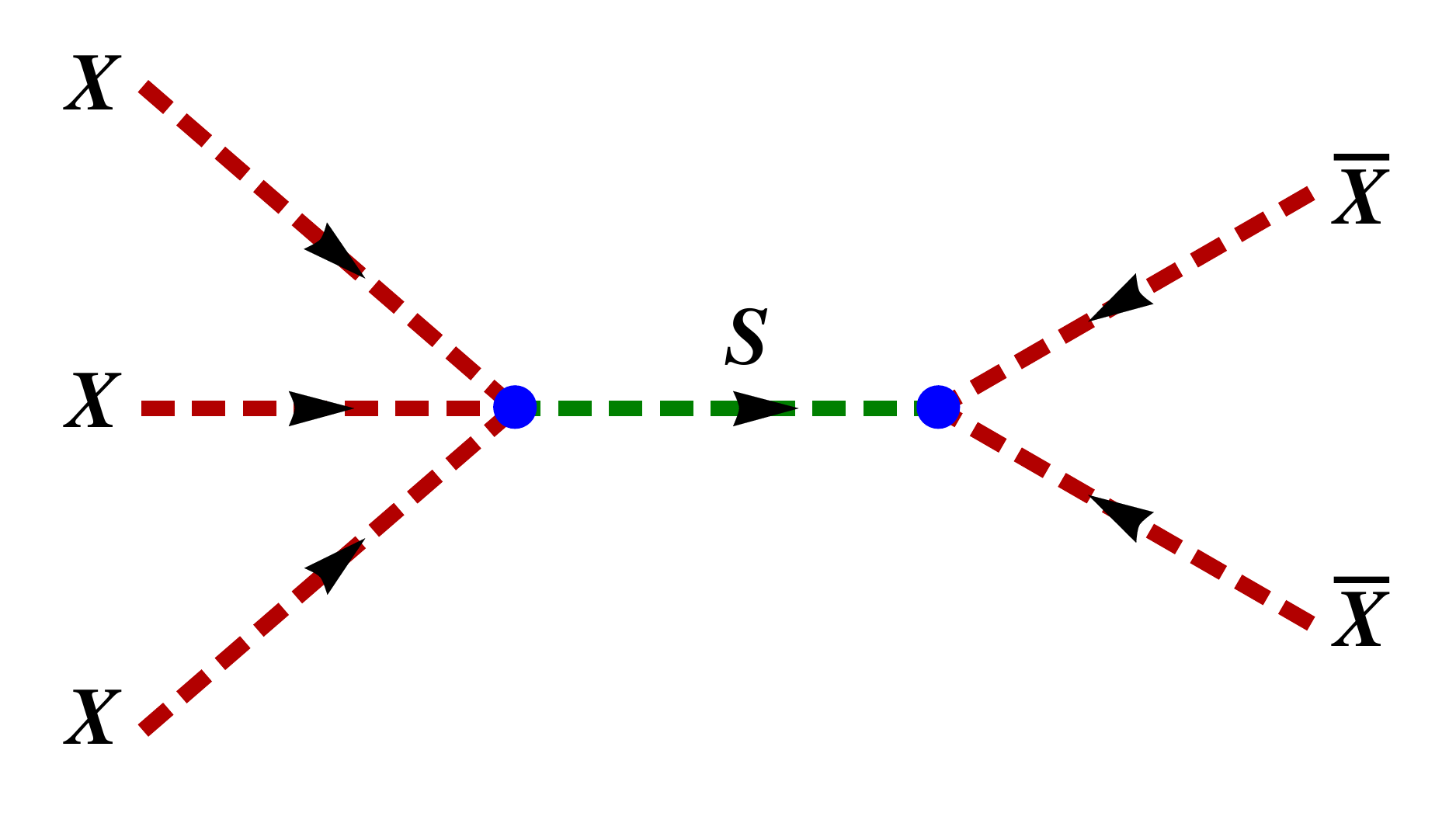}
\hspace{0.1cm}
\includegraphics[scale=0.25]{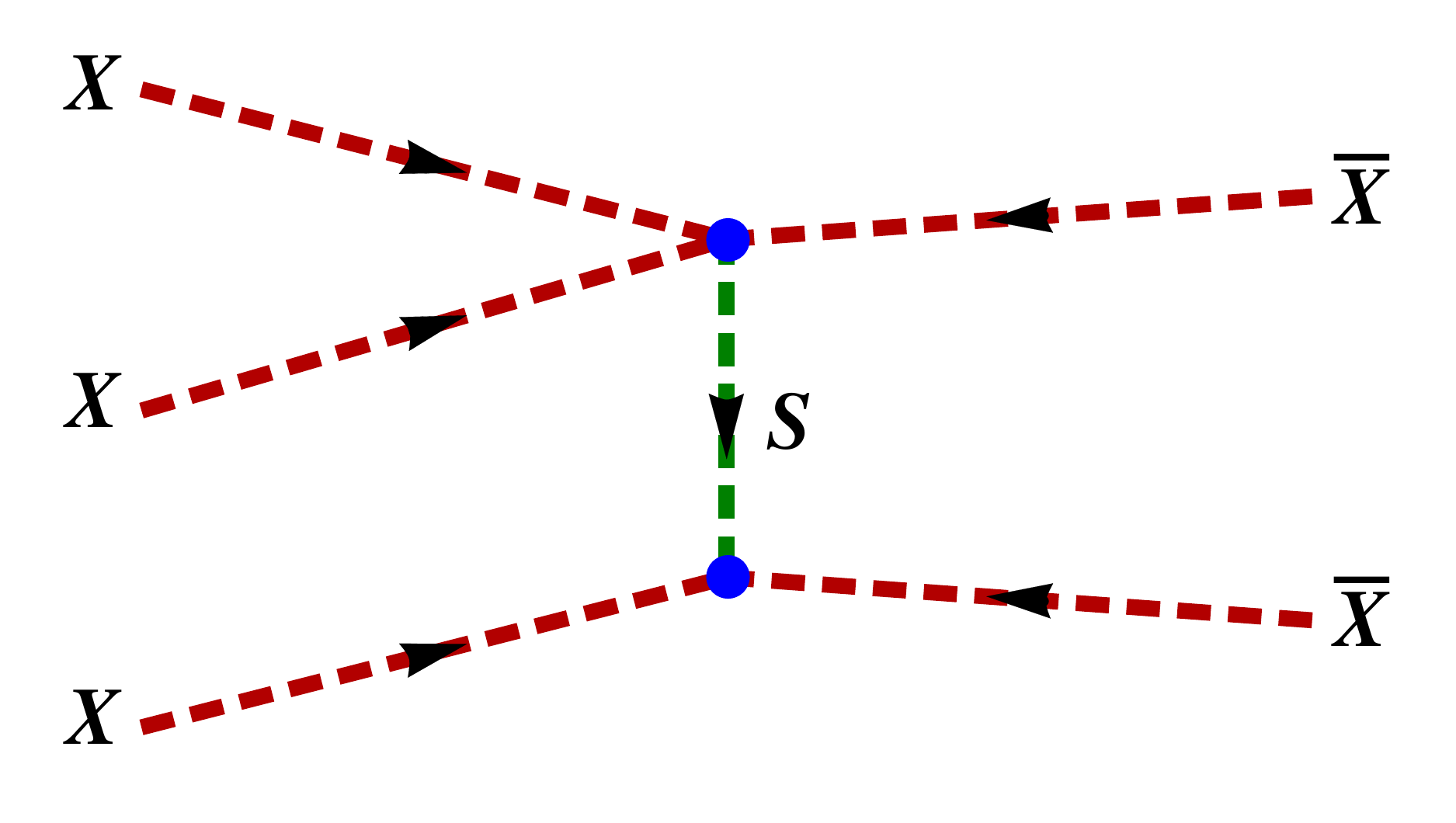}
\hspace{0.1cm}
\includegraphics[scale=0.25]{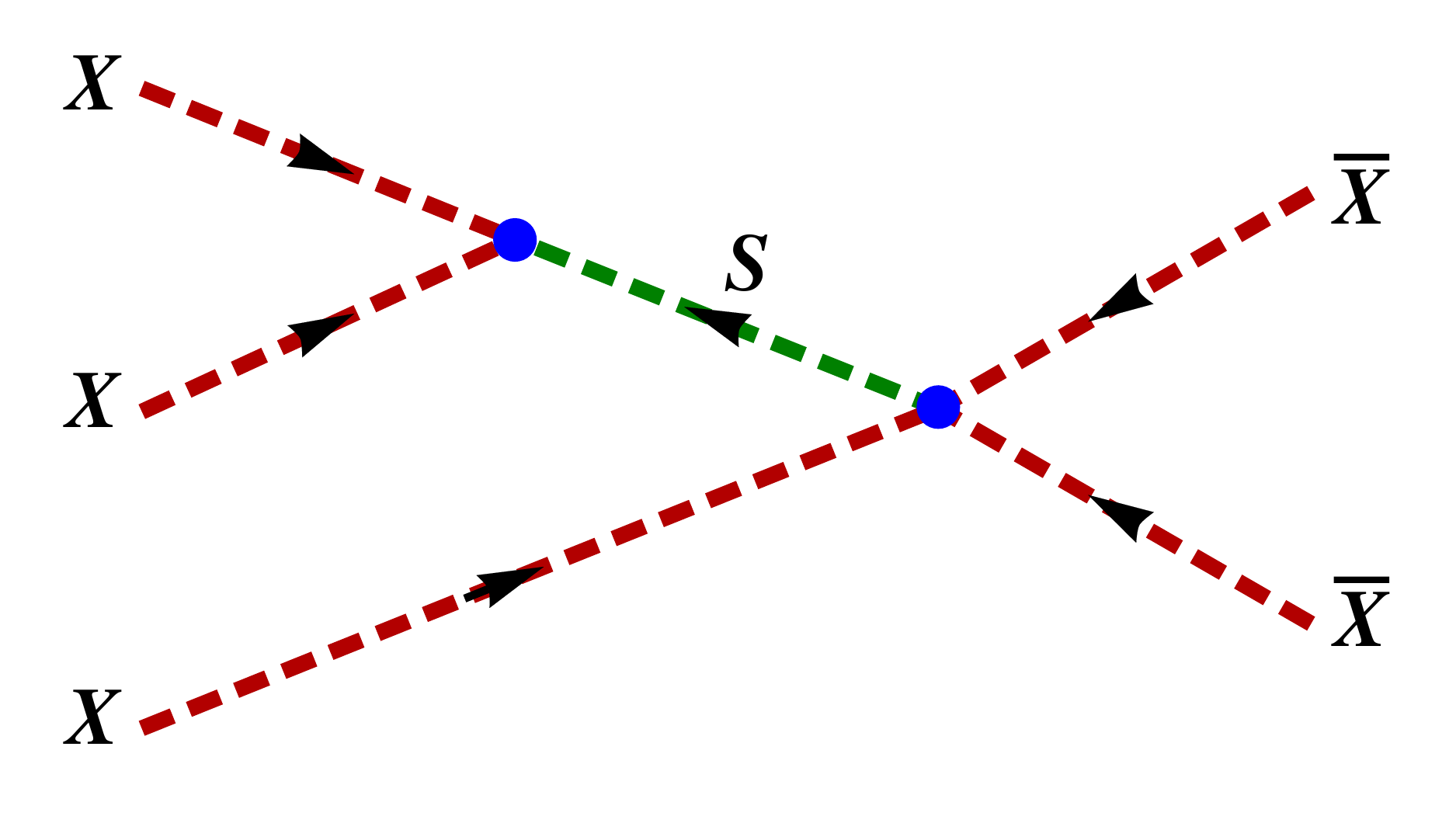} \\[0.1cm]
\hspace{0.02cm}
\includegraphics[scale=0.25]{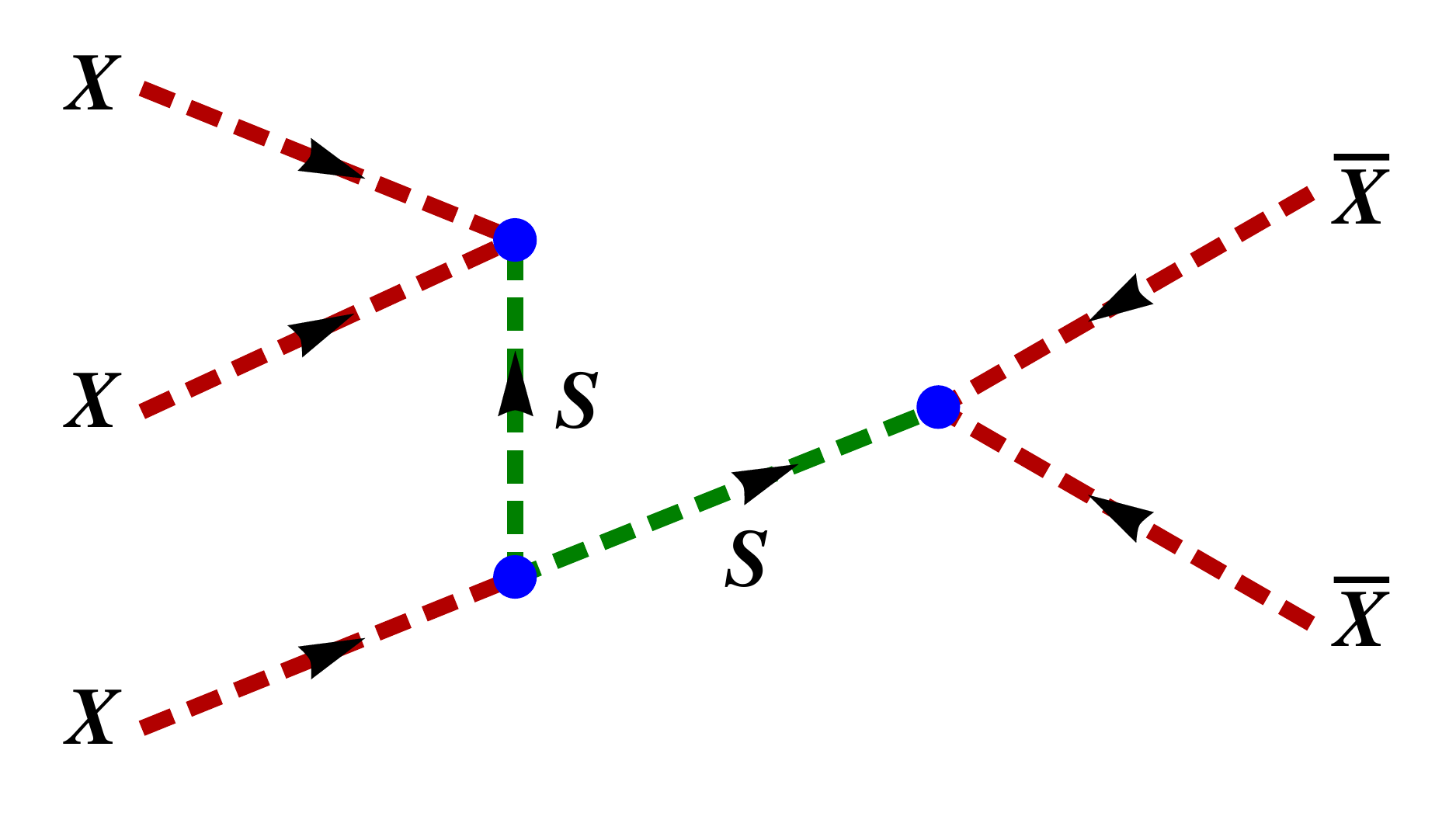}
\hspace{0.1cm}
\includegraphics[scale=0.25]{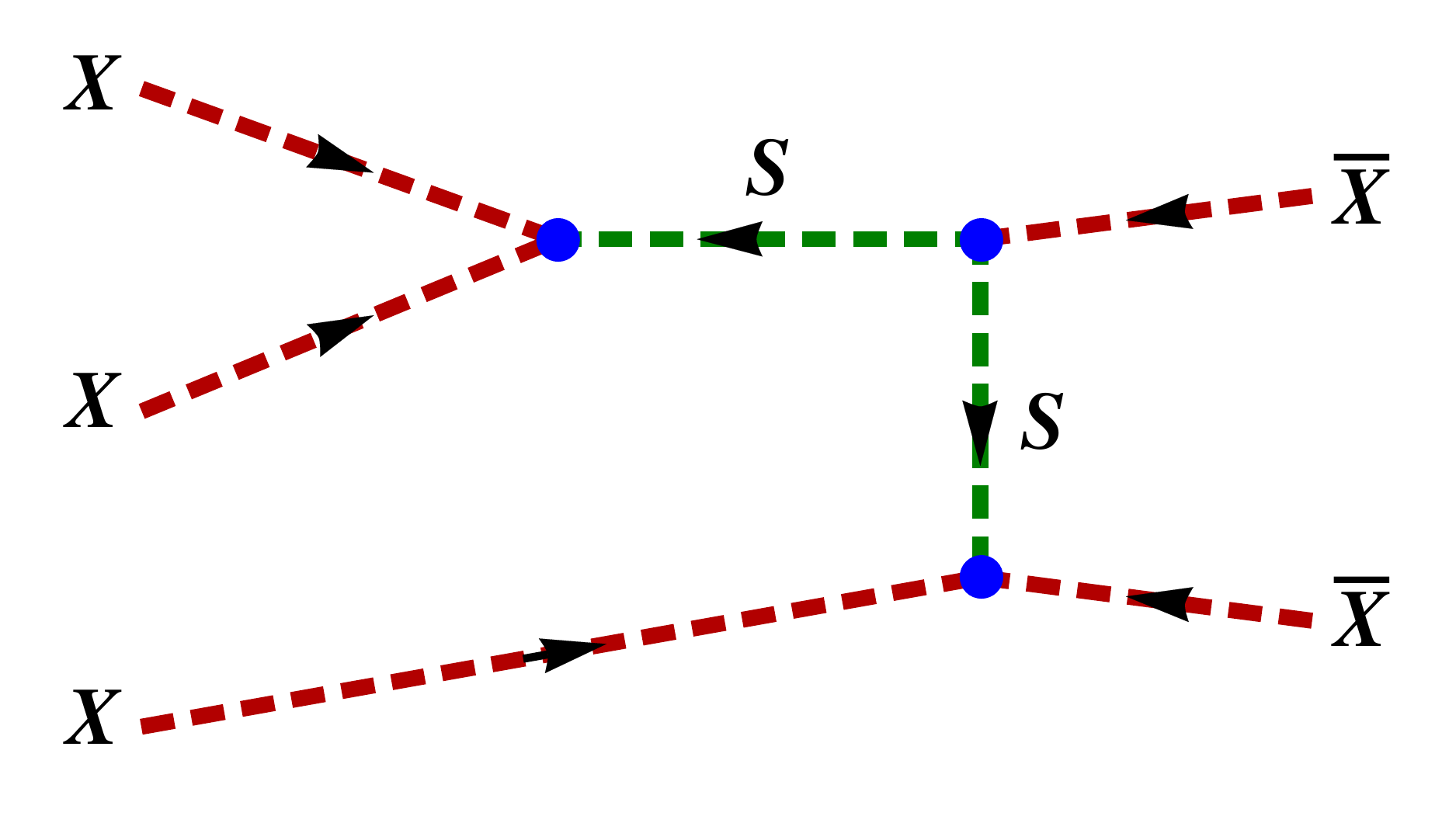}
\hspace{0.1cm}
\includegraphics[scale=0.25]{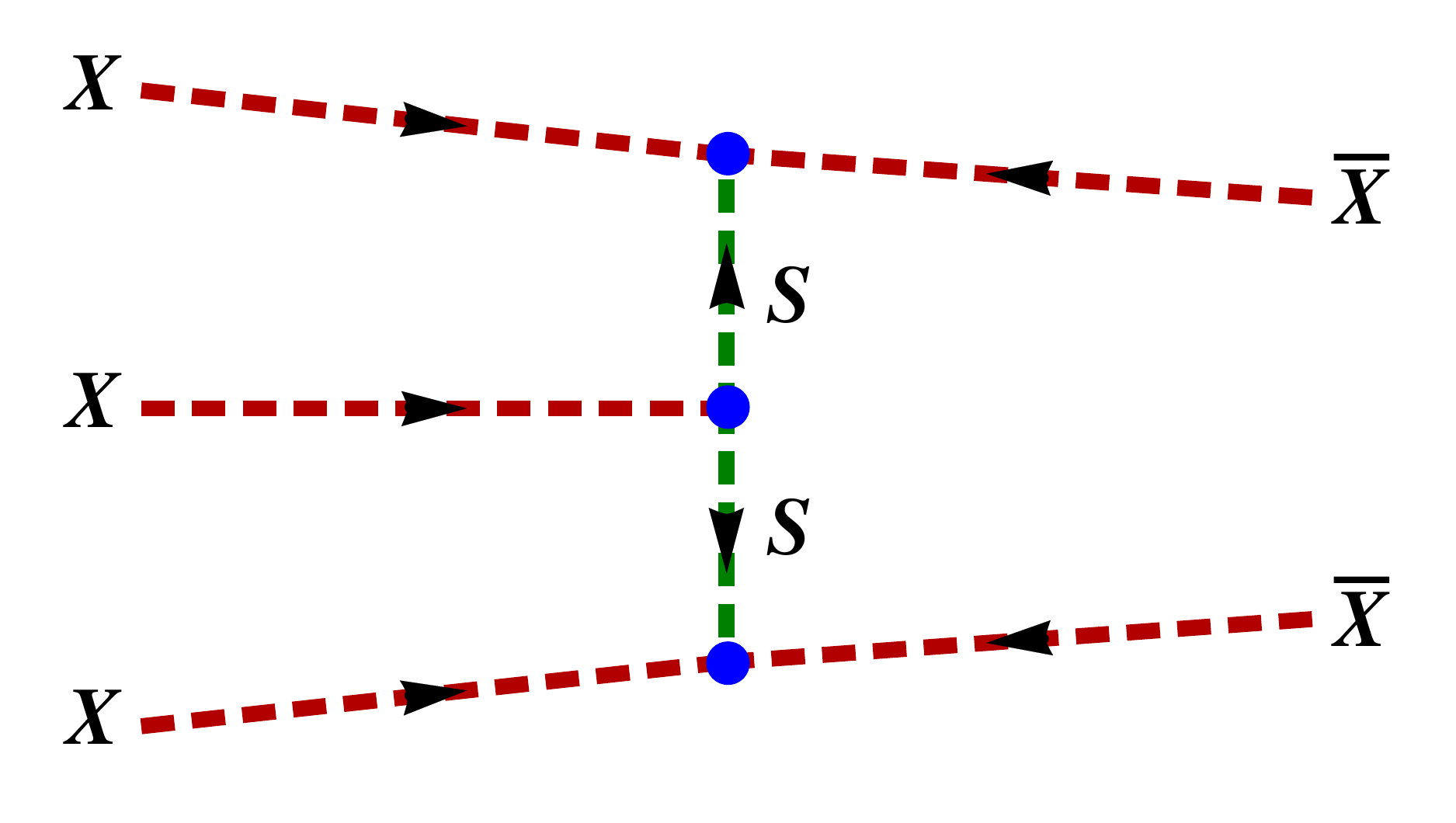}
\vspace{-0.3cm}
\caption{Feynman diagrams for the ${}^{}3{}^{} \to{}^{} 2{}^{}$ annihilation process $XXX \to \bar{X}\bar{X}$, where the similar diagrams obtained by crossing the contraction in the initial and the final states are not shown.}
\label{fig:3to2}
\end{center}
\vspace{-0.5cm}
\end{figure}
By applying the standard derivation~\cite{Kolb:1990vq}, the approximate solution to the Boltzmann equation for the current 
relic density ${}^{}\Omega_\text{DM}$ is given by
\begin{eqnarray}\label{relic}
\Omega_\text{DM}\hat{h}^2
\,\simeq\,
\frac{\,\,5.7\times10^8~\text{GeV}^{-1}}{m_X {}^{} g_{\star,f}^{3/4} {}^{} m_\text{pl}^{1/2} {}^{}  J^{1/2}} ~,\quad
J 
\,=\,  
\mathop{\mathlarger{\int_{x_f }^{{}^{}\infty}}} dx\, 
\frac{\big\langle \sigma_{3 \to 2}\upsilon_\text{rel}^2 {}^{} \big\rangle}{x^5} 
~,\quad
x_f \,\simeq \,20
~,\quad
\end{eqnarray}
where $x = m_X/T$, $\hat{h}$ denotes the normalized Hubble constant, $g_{\star,f}$ is the number of relativistic degrees of freedom at the freeze-out temperature, $T_f \,=\, m_X/x_f$, $m_\text{pl} = 1.22\times 10^{19}\,\text{GeV}$ is the Planck mass. To evaluate the thermal average of the $3 \to 2$ annihilation cross section, we employ the formula in Ref.~\cite{Choi:2017mkk} 
\begin{eqnarray}\label{thermal}
\big\langle \sigma_{3 \to 2}\upsilon_\text{rel}^2 {}^{} \big\rangle
\Eq 
\frac{x^3}{2}
\int_0^{{}^{}\infty} d\beta \, \big(\sigma_{3\to 2}\upsilon_\text{rel}^2\big) \, \beta^2 {}^{} {}^{}\text{exp}\big({-{}^{}x{}^{}\beta}{}^{}\big)
~,
\end{eqnarray}
where $\beta = \frac{1}{2}\big(\upsilon_1^2+\upsilon_2^2+\upsilon_3^2{}^{}\big)$ with $\upsilon_i$ the velocities of three initial DM particles. In the $\nu$SIMP model, the Feynman diagrams of the $3 \to 2$ process $XXX \to \bar{X}\bar{X}$ are shown in Fig.~\ref{fig:3to2}. From Eq.\eqref{3to2int}, the effective $3 \to 2$ annihilation cross section under CP invariance is calculated as
\vspace{0.1cm}
\begin{eqnarray}\label{sigma3to2}
\sigma_{3\to 2}\upsilon_\text{rel}^2
\Eq
\frac{\,25\sqrt{5}\,\mu_2^2 {}^{}{}^{} c_\xi^5\,}{9216 {}^{} \pi {}^{} m_X^3}
\scalebox{1.0}{\Bigg|}
\frac{3{}^{}\mu_1\mu_2\big(11 {}^{} m_X^4 - 8 {}^{} m_X^2m_S^2 + m_S^4\big)}
{\big(m_X^2 + m_S^2\big)^{\hspace{-0.05cm}2}\big(4 {}^{} m_X^2 - m_S^2 + i m_S\Gamma_S\big)
\big(\hat{s} - m_S^2 + i m_S\Gamma_S\big)}
\nonumber\\[0.1cm]
&&\hspace{2.2cm}-\,
\frac{\lambda_3\big(37 {}^{} m_X^4-21 {}^{} m_X^2m_S^2 + 2 {}^{} m_S^4\big)}
{\big(m_X^2 + m_S^2\big)\big(4 {}^{} m_X^2-m_S^2 + i m_S\Gamma_S\big)
\big(\hat{s} - m_S^2 + i m_S\Gamma_S\big)}
\scalebox{1.0}{\Bigg|}^2~,
\end{eqnarray}
where
$\hat{s}\,=(p_1+p_2+p_3)^2\,\simeq\,9{}^{}m_X^2\big(1+2{}^{}\beta/3\big)$
and the momenta of DM are neglected except around the resonance
$\hat{s}{}^{}{}^{}\approx m_S^2$. By taking the mass spectrum 
$2{}^{}M_k > m_S > 2{}^{}m_X$, the decay width of the particle $S$ is computed as
\begin{eqnarray}\label{GammaS}
\Gamma_S 
\,=\,
\Gamma\big(S\to X\bar{X}\big) 
\,=\,
\frac{\mu_2^2 {}^{}{}^{} c_\xi^2}{32{}^{}\pi{}^{}m_S}\sqrt{1-\frac{4{}^{}m_X^2}{m_S^2}} ~.
\end{eqnarray}
\vspace{-0.3cm}

To enhance the ${}^{}3{}^{} \to{}^{} 2{}^{}$ annihilation cross section, we pick the resonant pole $\,m_S {}^{}\simeq{}^{} \sqrt{\hat{s}} {}^{}\simeq {}^{}3{}^{}{}^{}m_X$ in Eq.\eqref{sigma3to2}, and it is convenient to adjust the resonant behavior by defining the following dimensionless parameters as
\begin{eqnarray}\label{resonant}
\epsilon_S \,=\, \frac{m_S^2-9{}^{}m_X^2}{9{}^{}m_X^2} ~,\quad
\gamma_S \,=\, \frac{m_S\Gamma_S}{9{}^{}m_X^2} 
~,
\end{eqnarray}
where ${}^{}\epsilon_S$ indicates the degeneracy between $m_S$ and
$3{}^{}m_X$, and $\gamma_S$ is the width of the resonance.\footnote{From
Eqs.\eqref{novev},  \eqref{GammaS} and \eqref{resonant} with $\xi \ll 1$
and $m_S \simeq 3{}^{}m_X$, one can easily show that $\gamma_S
\,\simeq{}^{} 10^{-3} {}^{} {\cal R}_2^2 \,\lesssim 10^{-2}
\lambda_X$. Thus one obtains $\gamma_S {}^{} \lesssim {}^{} 0.1{}^{} \ll
1$ with the perturbative bound $\lambda_X < 4{}^{}\pi$.}

\begin{figure}[t]
\begin{center}
\includegraphics[scale=0.54]{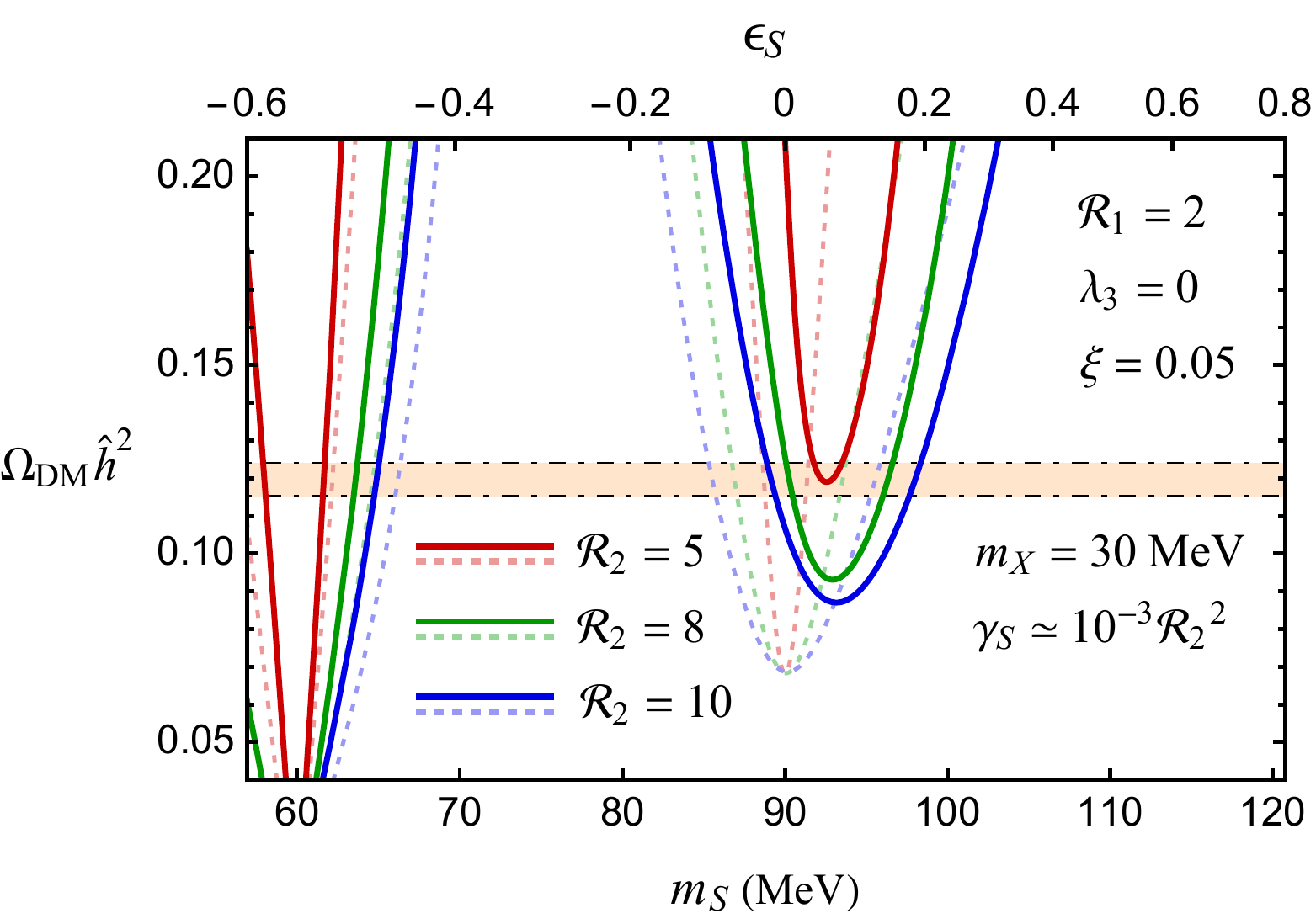}
\hspace{0.05cm}
\includegraphics[scale=0.54]{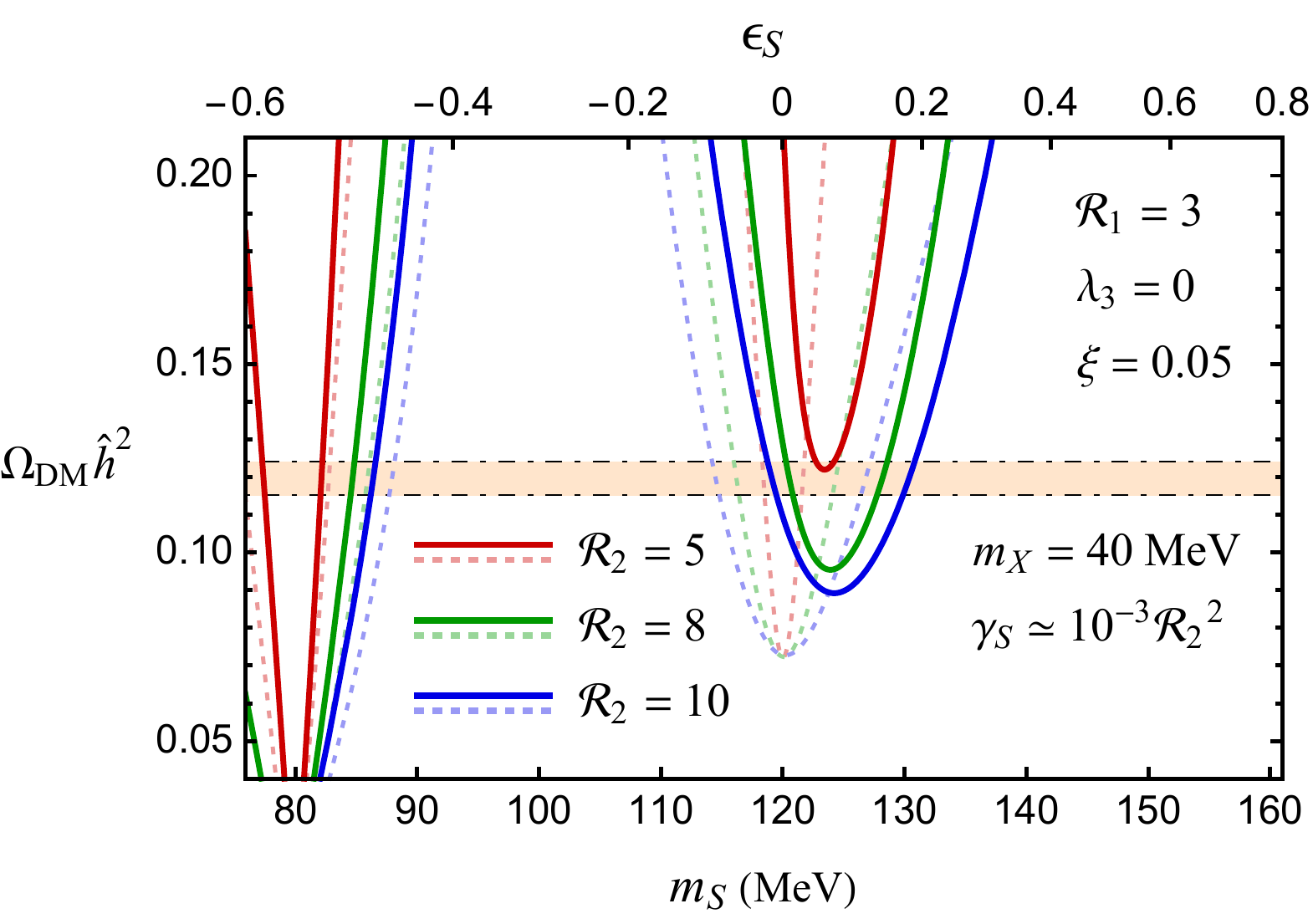}
\vspace{-0.3cm}
\caption{The predicted relic density versus $m_S$ for nonzero (zero) temperature of DM in solid lines (light dashed lines). The orange band is the observed value ${}^{}0.1153 \leq \Omega_\text{DM}\hat{h}^2 \leq 0.1241{}^{}$ at the 95\%\,C.L..}
\label{fig:Relic}
\end{center}
\vspace{-0.6cm}
\end{figure}

With these variables, the ${}^{}3{}^{} \to{}^{} 2{}^{}$ annihilation cross section can be expressed 
in the Breit-Wigner resonant form similar to the one in Ref.\cite{Gondolo:1990dk}
\begin{eqnarray}\label{BreitWigner}
\sigma_{3\to 2} \upsilon_\text{rel}^2 
\,=\,
\frac{c_X^{}}{m_X^5}
\frac{\gamma_S^2}{\big(\epsilon_S - 2 {}^{} \beta/3\big)^{\hspace{-0.05cm}2} + \gamma_S^2}~,
\end{eqnarray}
where the coefficient $c_X^{}$ is 
\begin{eqnarray}
c_X^{} \Eq
\frac{25\sqrt{5}\,\pi{}^{}c_\xi{}^{}m_S^2}
{\big(m_S^2-4 {}^{} m_X^2\big)
\scalebox{1.1}{\big[}
\big(m_S^2-4 {}^{} m_X^2\big)^{\hspace{-0.06cm}2}+m_S^2\Gamma_S^2
\scalebox{1.1}{\big]}
\big(m_S^2+m_X^2\big)^{\hspace{-0.06cm}2}}
\nonumber\\[0.1cm]
&&\times
\Bigg[
\frac{{\cal R}_1 {}^{} m_X^2 \big(m_S^4-8{}^{}m_S^2m_X^2 +11{}^{}m_X^4\big)}{m_S^2+m_X^2} -
\frac{\lambda_3 \big(2 {}^{} m_S^4-21{}^{} m_S^2 m_X^2 + 37 {}^{} m_X^4\big)}{3 {}^{} {\cal R}_2}
\Bigg]^{\hspace{-0.07cm}2}
~,
\end{eqnarray}
with ${\cal R}_{1,2} = \mu_{1,2}/m_X$. Utilizing Eq.\eqref{relic}, we
present the plots of the DM relic density
${}^{}\Omega_\text{DM}{}^{}$\,versus $m_S$ with different values of
$\,m_X$ and ${}^{}{\cal R}_{1,2}$ in Fig.~\ref{fig:Relic}, where the
solid lines (light dashed lines) are the predicted values by using the
thermal (non-thermal) averaged effective 
$3 \to 2$ annihilation cross section. The orange region is the latest relic density data $\,\Omega_\text{DM}\hat{h}^2 = 0.1197 
\pm 0.0022\,$ given by the Planck Collaboration~\cite{Ade:2015xua}. In these plots, we do not vary the mixing angle 
${}^{}\xi{}^{}$ since dependence of the mixing angle is extremely small as long as
$s_\xi\ll1$. Also, in order to examine the conditions of ${}^{}\langle X
\rangle = 0$ in Eq.\eqref{novev} easily,  we again assume $\lambda_3 =
0$. For nonzero $\lambda_3{}^{}$, the numerical results are similar as
pointed out in Ref.~\cite{Choi:2016tkj}. We have checked our choices of
the values of $\,{\cal R}_{1,2}$ can accommodate the requirements of
perturbativity, ${\cal R}_1^2 < \lambda_S < 4{}^{}\pi$ and ${\cal
R}_2^2/9 \lesssim \lambda_X  < 4{}^{}\pi$ with $m_S \simeq
3{}^{}m_X$. According to the plots, one can see that the low values of
${}^{}{\cal R}_{1,2}$ are disfavored if the DM mass $m_X$ is heavier.

\begin{figure}[b]
\vspace{0.1cm}
\begin{center}
\includegraphics[scale=0.21]{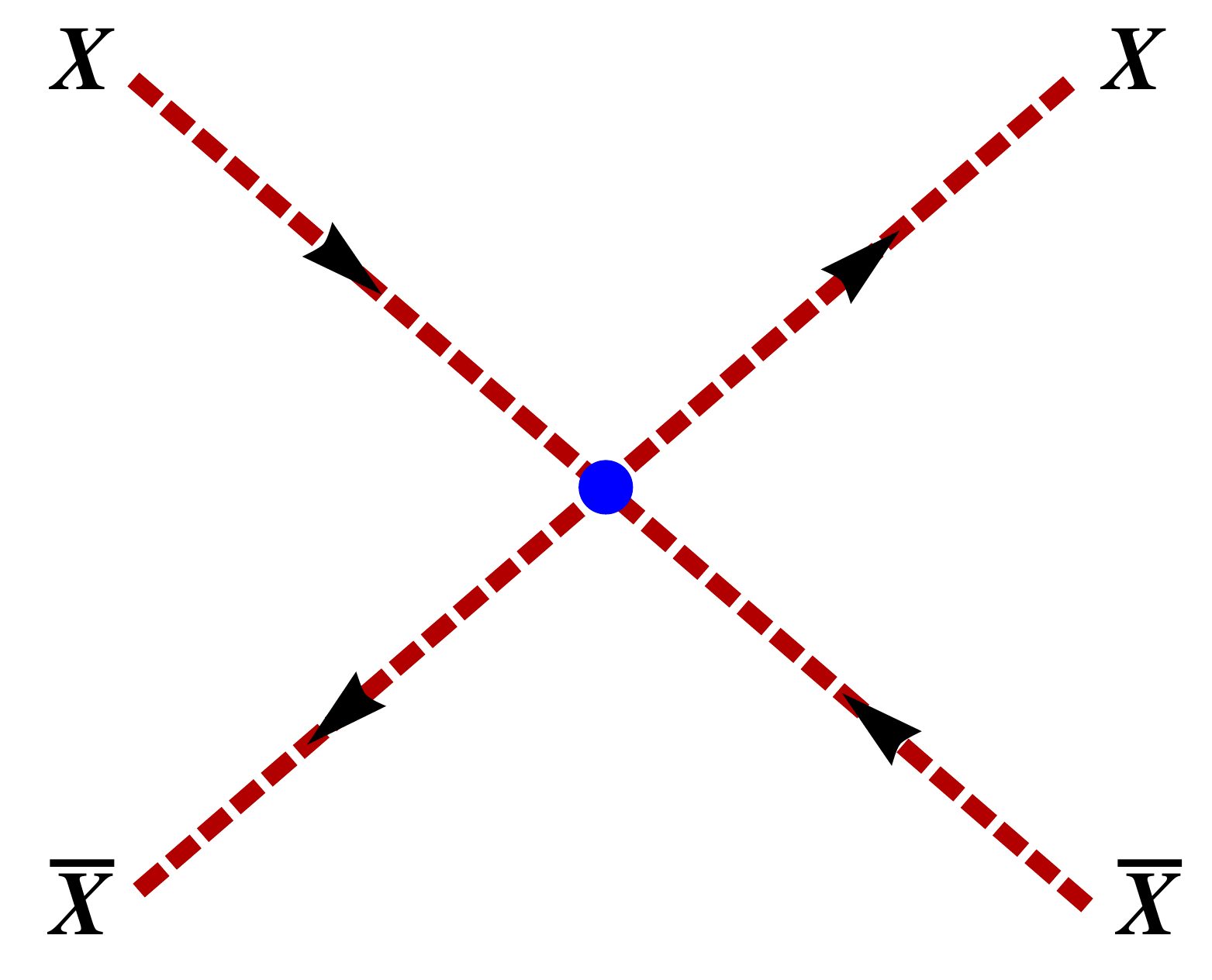}
\hspace{0.3cm}
\includegraphics[scale=0.21]{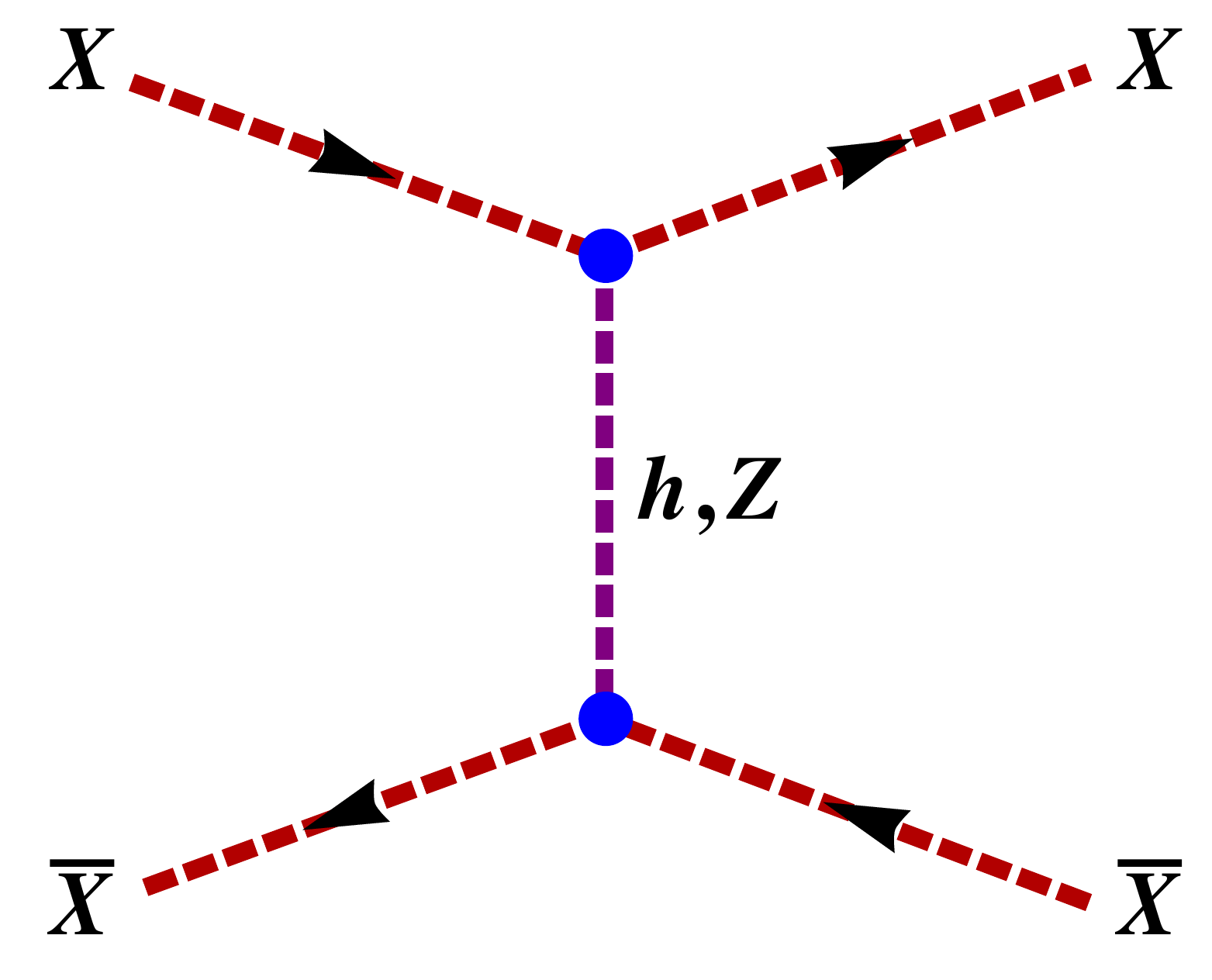}
\hspace{0.3cm}
\includegraphics[scale=0.21]{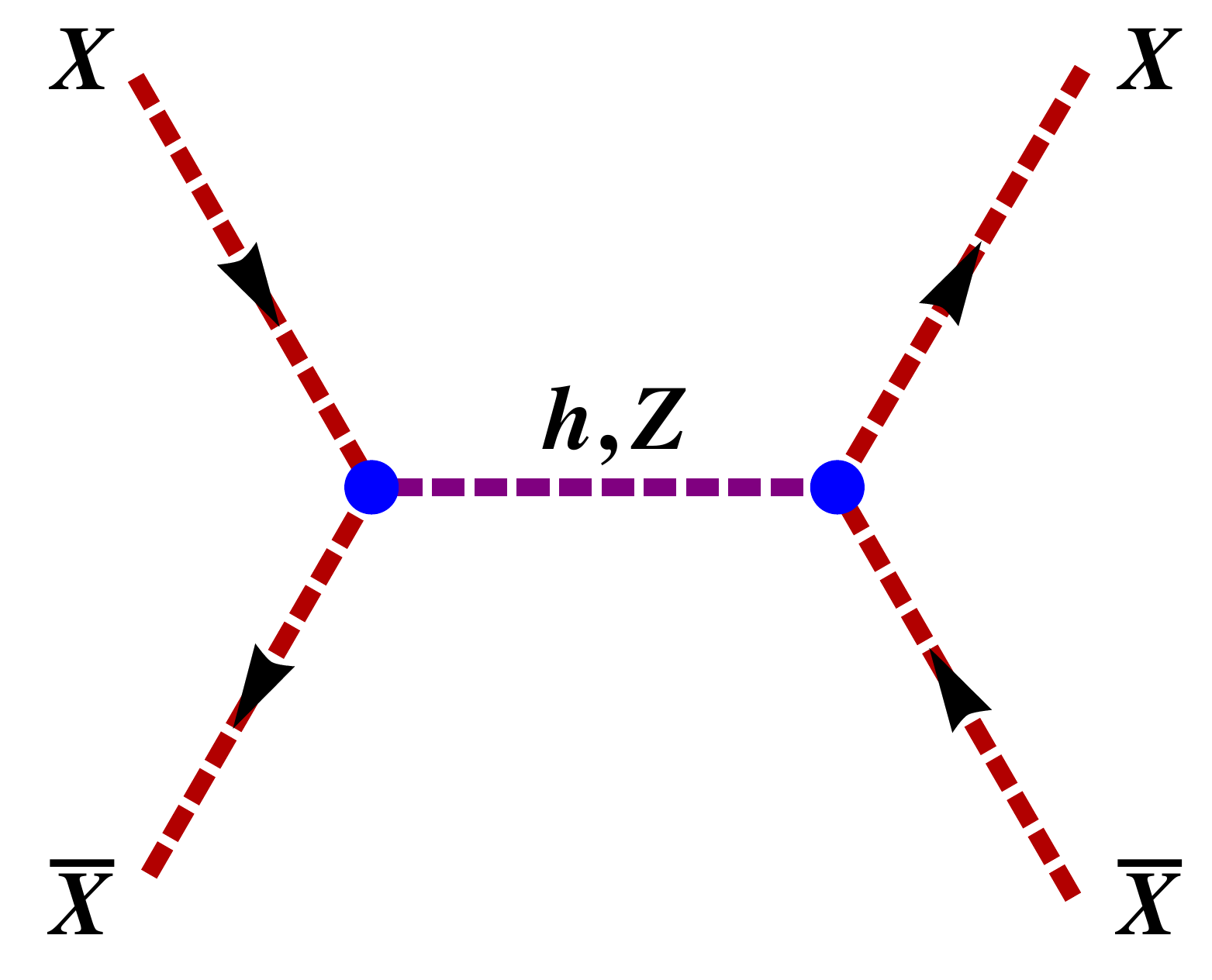}
\hspace{0.3cm}
\includegraphics[scale=0.21]{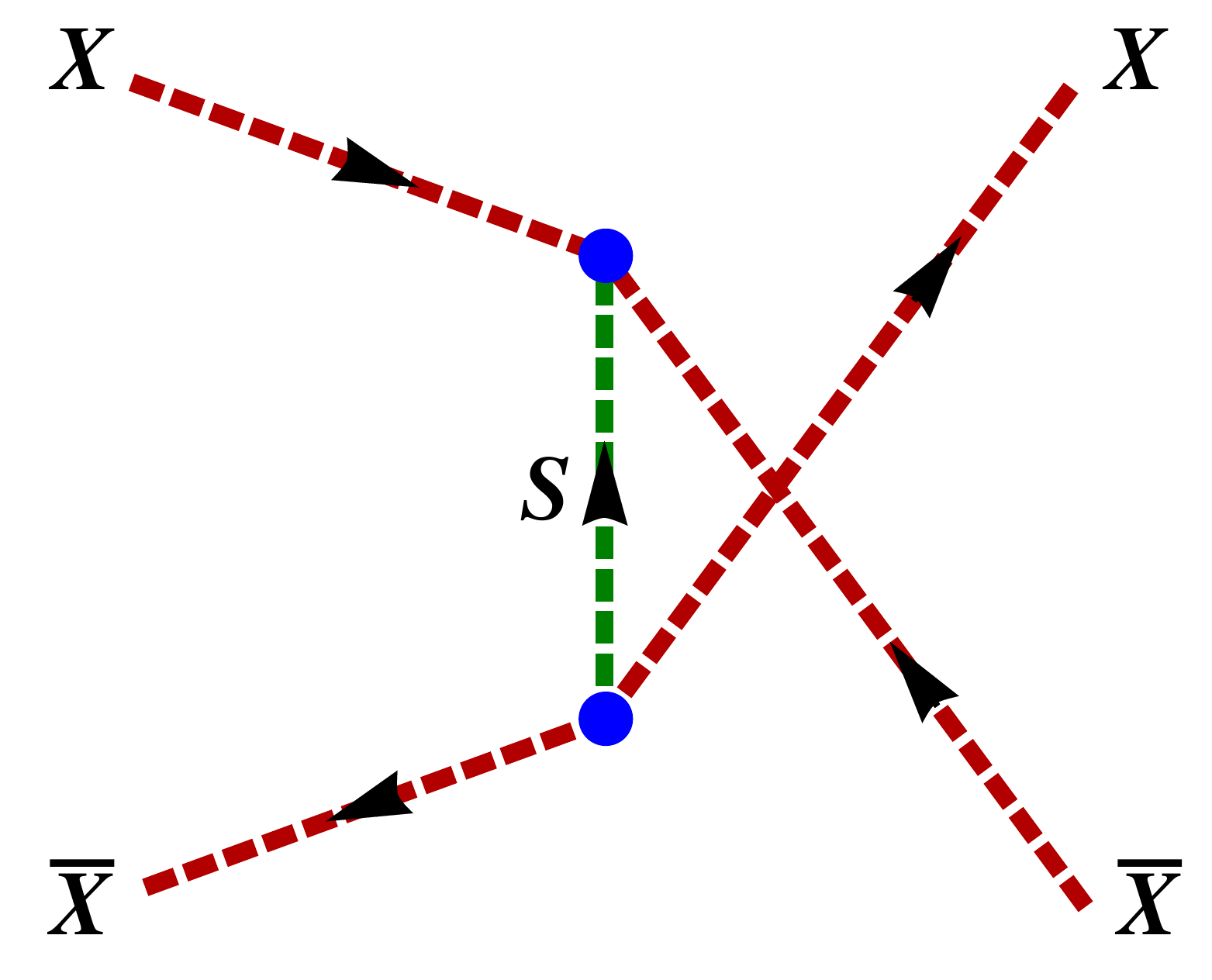}
\\[0.3cm]
\includegraphics[scale=0.21]{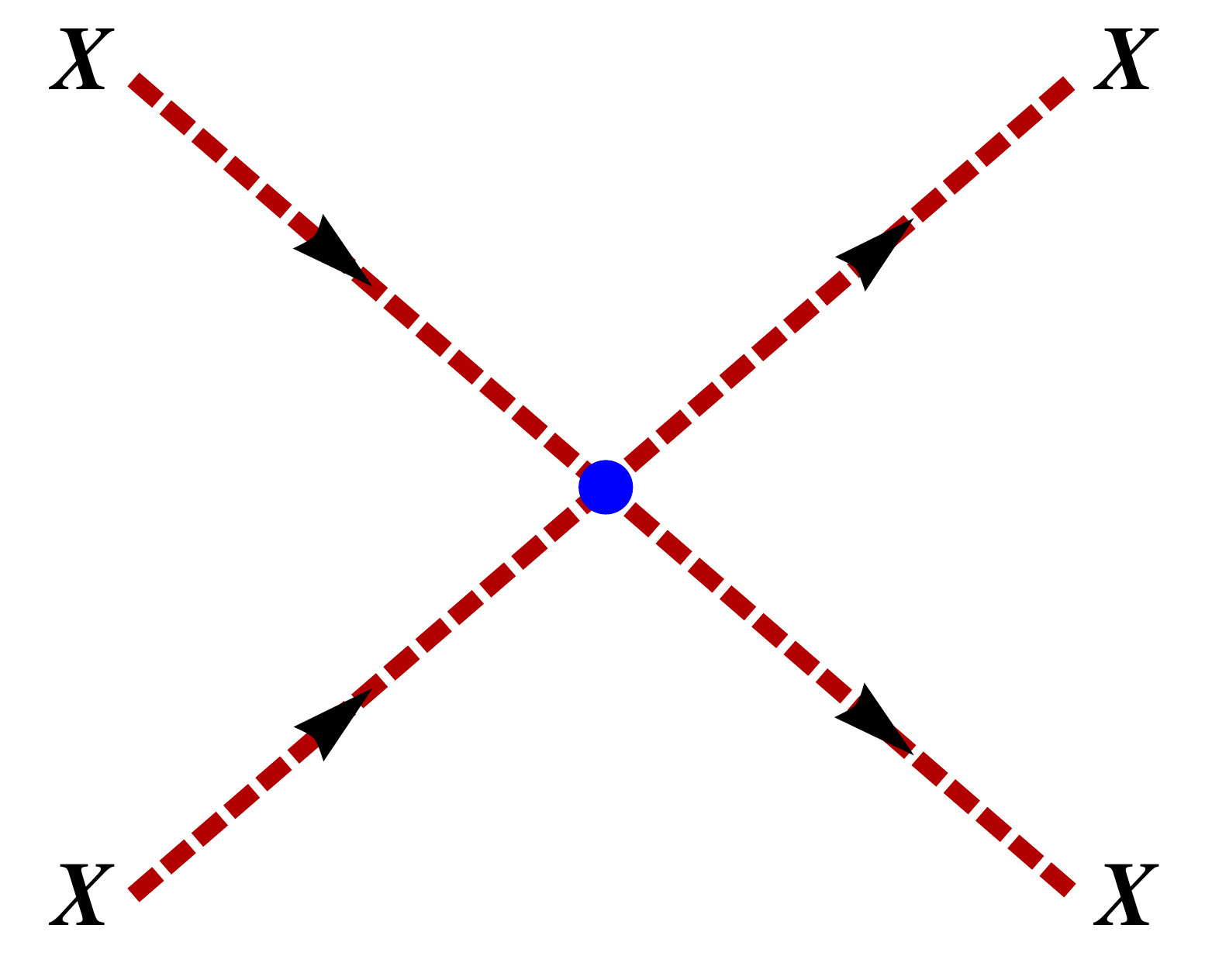}
\hspace{0.3cm}
\includegraphics[scale=0.21]{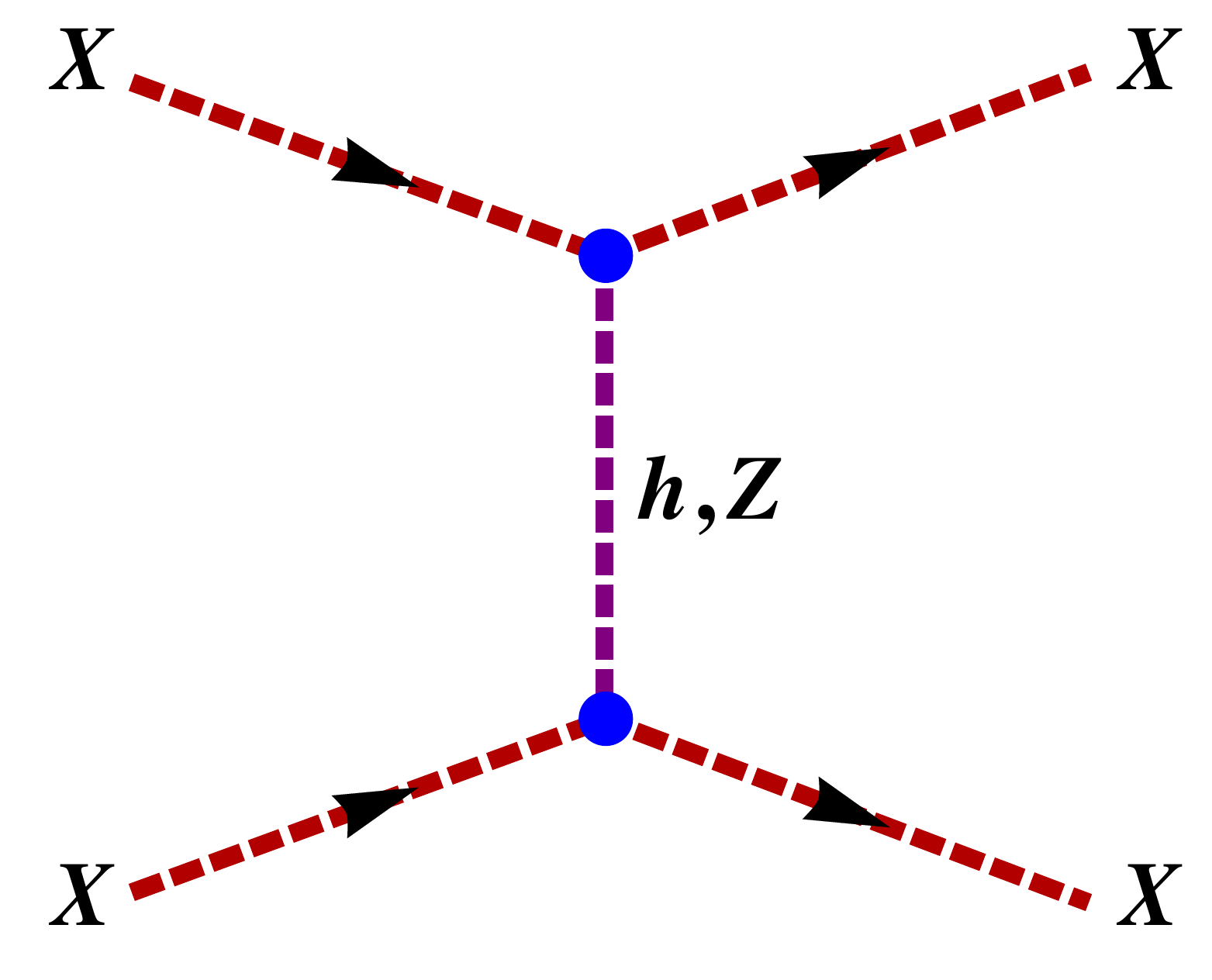}
\hspace{0.3cm}
\includegraphics[scale=0.21]{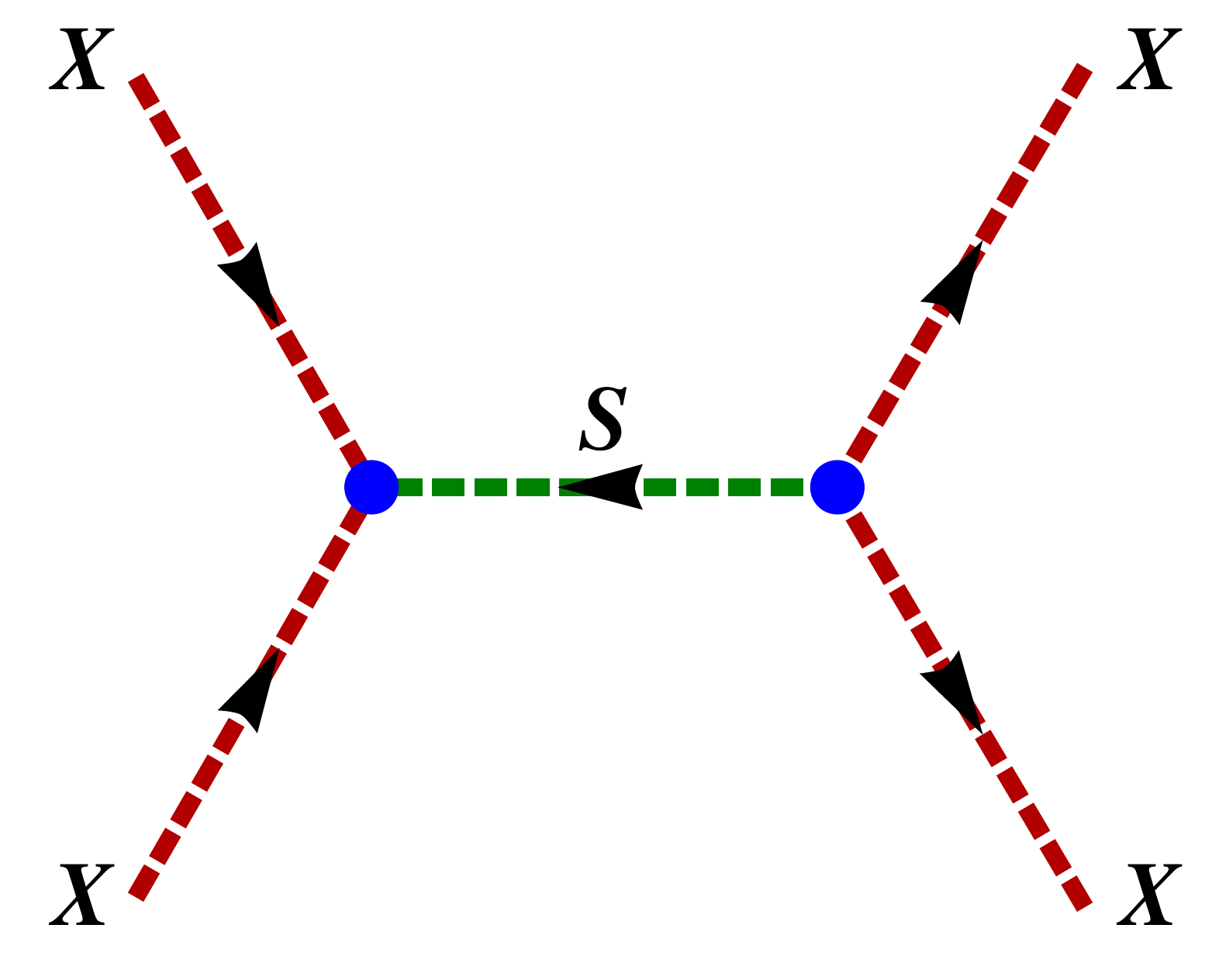}
\vspace{-0.1cm}
\caption{Feynman diagrams for the DM self-interacting processes. The upper (lower) diagrams correspond to the process 
$X\bar{X} \to X\bar{X} \, \big(XX \to XX {}^{} \big)$. For the process $\bar{X}\bar{X} \to \bar{X}\bar{X}$, the relevant diagrams can be obtained by flipping the arrows in the lower ones.}
\label{fig:self}
\end{center}
\end{figure}

Besides fitting the relic abundance of DM, there are the other astrophysical observations from the Bullet cluster~\cite{Markevitch:2003at,Clowe:2003tk,Randall:2007ph} and 
spherical halo shapes~\cite{Peter:2012jh}, which impose the bound ${}^{}\sigma_\text{self}/m_X {}^{} \lesssim {}^{} 1\,\text{cm}^2/\text{g}{}^{}$ with ${}^{}\sigma_\text{self} ={}^{} \frac{1}{4}(\sigma_{XX \to XX}+\sigma_{X\bar{X} \to X\bar{X}}+\sigma_{\bar{X}\bar{X} \to \bar{X}\bar{X}}){}^{}$ the effective self-interacting cross section. We depict in Fig.~\ref{fig:self} the Feynman diagrams of 
the DM self-interacting processes in our SIMP\,${}^{}$model, and their cross sections are calculated as
\begin{eqnarray}
\sigma_{X\bar{X}\to X\bar{X}} 
\Eq
\frac{1}{64^{}\pi^{}m_X^2}
\scalebox{1.1}{\bigg(}
\hspace{-0.05cm}
\lambda_X -
\frac{m_X^2}{m_S^2} {}^{} {\cal R}_2^2 {}^{} c_\xi^2
\scalebox{1.1}{\bigg)}^{\hspace{-0.15cm}2}
~,
\nonumber\\[0.1cm]
\sigma_{XX \to XX} 
\,=\, 
\sigma_{\bar{X}\bar{X}\to \bar{X}\bar{X}} 
\Eq
\frac{1}{128 {}^{} \pi {}^{} m_X^2}
\scalebox{1.1}{\bigg(}
\hspace{-0.05cm}
\lambda_X +
\frac{m_X^2}{4 {}^{} m_X^2 - m_S^2} {}^{} {\cal R}_2^2 {}^{} c_\xi^2
\scalebox{1.1}{\bigg)}^{\hspace{-0.15cm}2}
~,\quad
\end{eqnarray}
here we have neglected the contributions from the $h$ and $Z$-mediated diagrams due to their small couplings and mass suppression. By choosing an appropriate value of ${}^{}\lambda_X {}^{} \big({}^{}{\cal R}_2^2{}^{}m_X^2/m_S^2 < \lambda_X < 4{}^{}\pi\big)$, the bounds from the Bullet cluster and spherical halo shapes can be satisfied. For instance, if we take $m_X{}^{}(m_S) = 30\,(93)\,\text{MeV}, {\cal R}_{1,2} = 2,5{}^{}$ and ${}^{}\xi =0.05$ with ${}^{}\lambda_X \hspace{-0.02cm} = 7$, we find $\sigma_\text{self}/m_X \simeq {}^{} 0.26 \,\text{cm}^2/\text{g}{}^{}$. More examples and discussions can be found in Ref.~\cite{Choi:2016tkj}.

\section{SIMP condition}\label{sec:6}
In the SIMP$\,{}^{}$paradigm, DM is thermally produced through the
${}^{}3 \to 2{}^{}$ annihilation process into the particles in the dark
sector rather than the ${}^{}2 \to 2{}^{}$ annihilation process into the
SM$\,{}^{}$particles. 
On the other hand, in order to keep the temperature of the dark sector as the same
with the SM sector, the SIMP$\,{}^{}{}^{}$candidate needs to be in
kinetic equilibrium with the SM sector.
Thus the naive criteria that DM can be a SIMP$\,{}^{}$candidate is given by~\cite{Hochberg:2014dra}
\begin{eqnarray}\label{SIMPcondition}
\Gamma_{2 \to 2}  \,<\,
\Gamma_{3 \to 2}  \,<\,
\Gamma_\text{kin}  ~,
\end{eqnarray}
which should be held during the freeze-out temperature. In this inequality, each reaction rate is defined by ${}^{}\Gamma_{2 \to 2} = n_X \langle \sigma_{2 \to 2} \upsilon_\text{rel} \rangle{}^{}, \, \Gamma_{3 \to 2} = n_X^2 \langle \sigma_{3 \to 2} \upsilon_\text{rel}^2 \rangle{}^{}$, and ${}^{}\Gamma_\text{kin} = n_\text{SM} \langle \sigma_\text{kin} \upsilon_\text{rel} \rangle{}^{}$,\footnote{In the WIMP paradigm, the Boltzmann equation of the DM number density is given by
\vspace{-0.15cm}
\begin{equation}
\dot{n}_\text{DM} + 3 {}^{}{}^{} {\cal H} {}^{} n_\text{DM} = 
- \,\langle \sigma_{2 \to 2}\upsilon_\text{rel} {}^{} \rangle 
\scalebox{1.2}{\big[} n_\text{DM}^2-(n_\text{DM}^\text{eq}\big){}^{\hspace{-0.05cm}2} {}^{} \scalebox{1.2}{\big]} ~,
\nonumber\\[-0.15cm]
\end{equation}
where $\langle \sigma_{2 \to 2}\upsilon_\text{rel} {}^{}\rangle$ is the
thermal averaged effective $2 \to 2$ annihilation cross section. Due to
this definition, an extra factor $1/2$ is multiplied to the DM cross
sections (Eq.(\ref{density}) and (\ref{eq:ann})). This comes from the
fact that DM and anti-DM particles are not identical in our case~\cite{Gondolo:1990dk}. 
\vspace{0.1cm}} 
where the number densities of DM and the SM particles are given as~\cite{Lee:2015gsa}
\begin{eqnarray}\label{densityX}
n_X {}^{}{}^{}=\, n_{\bar{X}} 
\,\simeq\, 
\frac{2.04 \times 10^{-9}\,\text{GeV}}{m_X} \, T^3 ~ , \quad
n_\text{SM} 
\,=\, 
\frac{g}{2{}^{}\pi^2} {}^{}{}^{} T^3  \hspace{-0.1cm}
\mathop{\mathlarger{\int}_0^{{}^{}\infty}}\hspace{-0.1cm} \frac{z^2 {}^{} dz}
{\text{exp}\scalebox{1.1}{\big[}{\sqrt{z^2+(m_\text{SM}/T)^2}}\,\scalebox{1.1}{\big]}\pm 1} ~, 
\end{eqnarray}
with ${}^{}g{}^{}$ counts the internal degrees of freedom, $m_\text{SM}$
being the mass of the SM particle, $(+)$ applies to fermions, and $(-)$
pertains to bosons. 
It is also pointed out in Ref.~\cite{Kuflik:2015isi} that a slightly stronger SIMP condition may be derived by considering the
rate of energy transfer (rather than the rate of reaction) between the SM and dark sectors. This rigorous SIMP condition can be 
written as $|{}^{}\dot{E}_{3 \to 2}| {}^{}<{}^{} |{}^{}\dot{E}_\text{kin}|$, where
${}^{}\dot{E}_{3 \to 2}{}^{}$ is the rate of the DM mass transferring to the kinetic energy in the DM bath, and ${}^{}\dot{E}_\text{kin}{}^{}$ is the rate of the kinetic energy of DM transfers to the thermal plasma. Quoting the detailed calculations of the SIMP condition in Ref.~\cite{Hochberg:2015vrg}, we then impose $\Gamma_{3 \to 2} < \Gamma_{3 \to 2} < 10^{-2} {}^{} \Gamma_\text{kin}$ in our numerical study.

\begin{figure}[t]
\begin{center}
\includegraphics[scale=0.18]{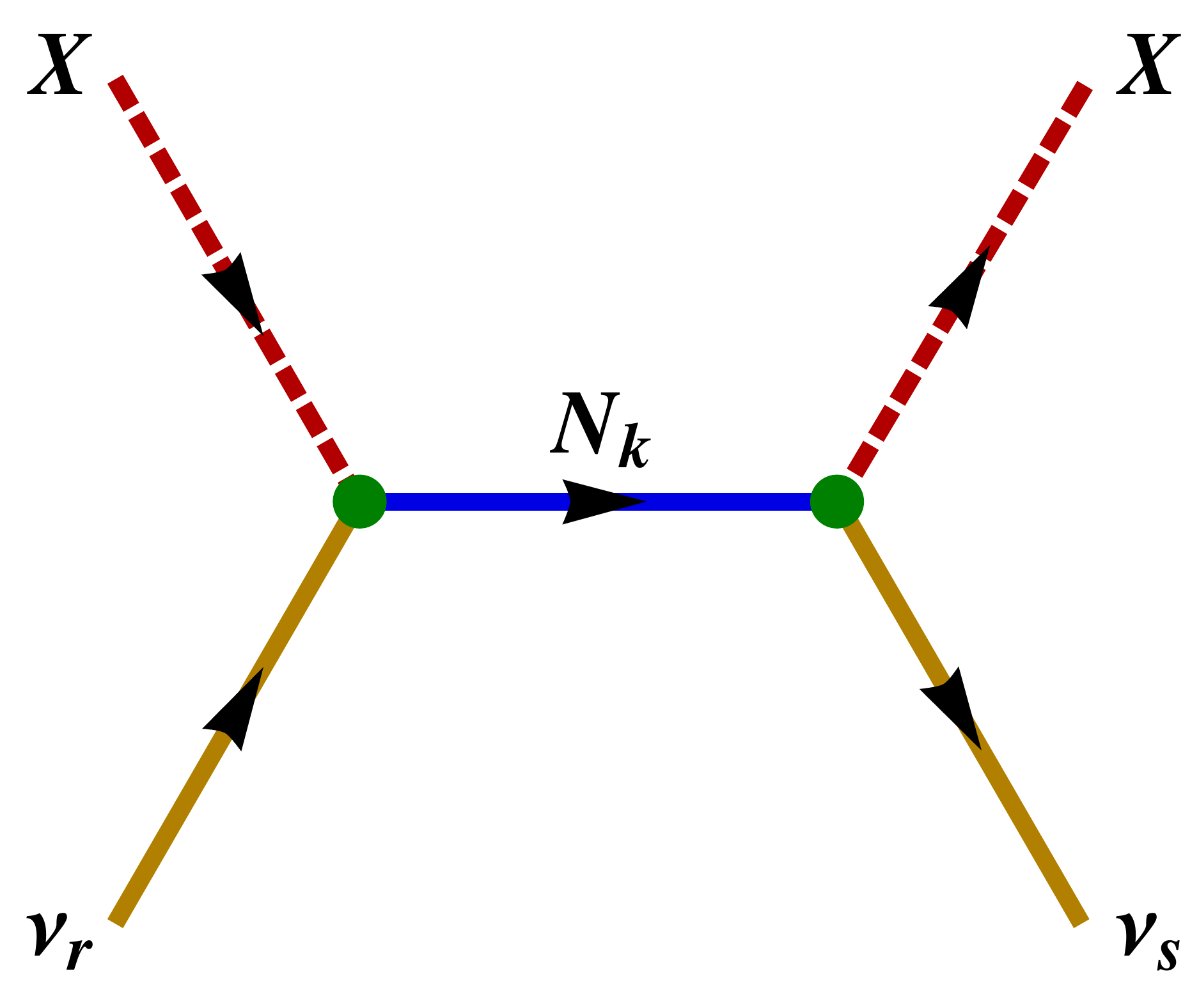}
\hspace{0.3cm}
\includegraphics[scale=0.18]{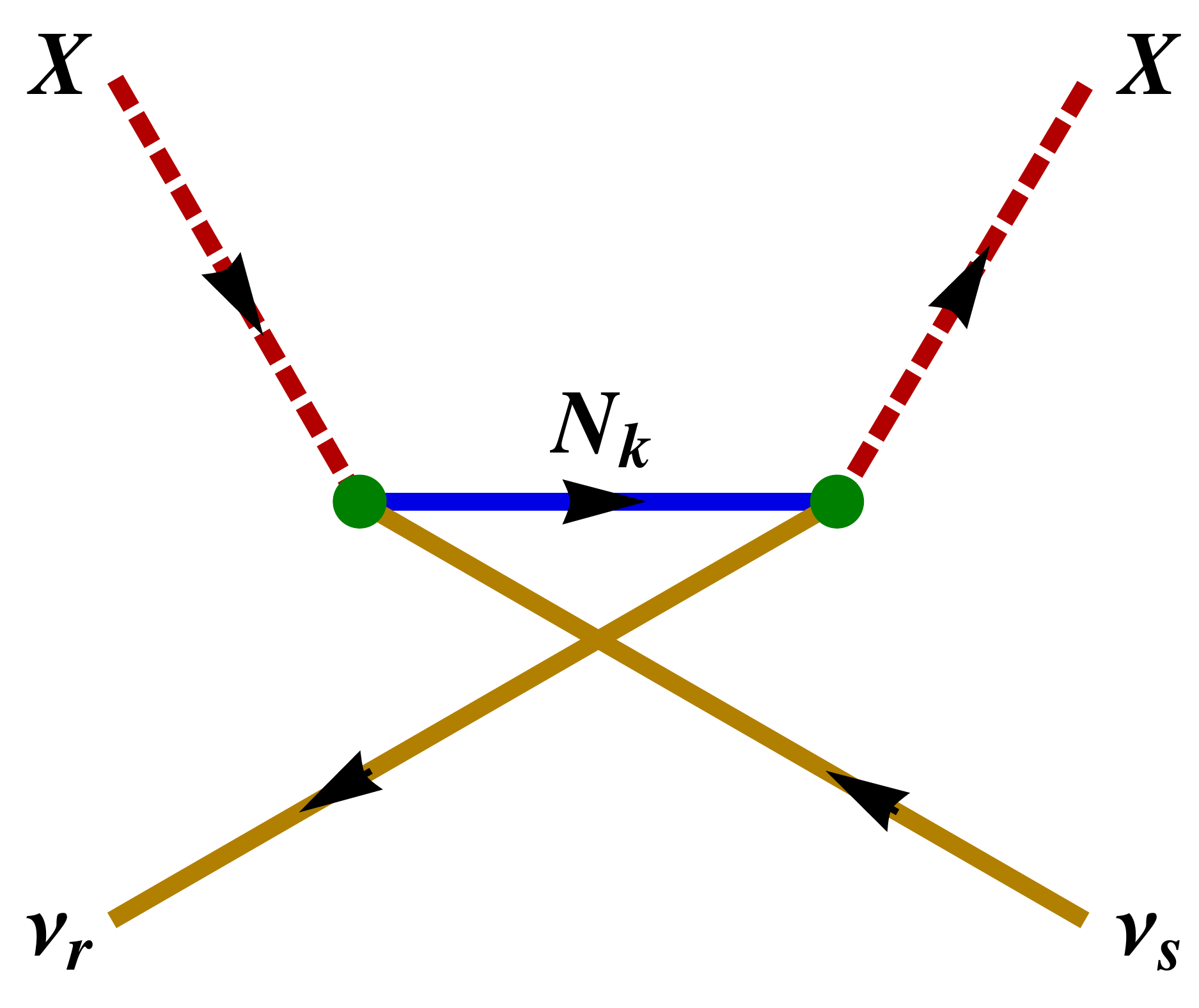}  
\hspace{0.3cm}
\includegraphics[scale=0.18]{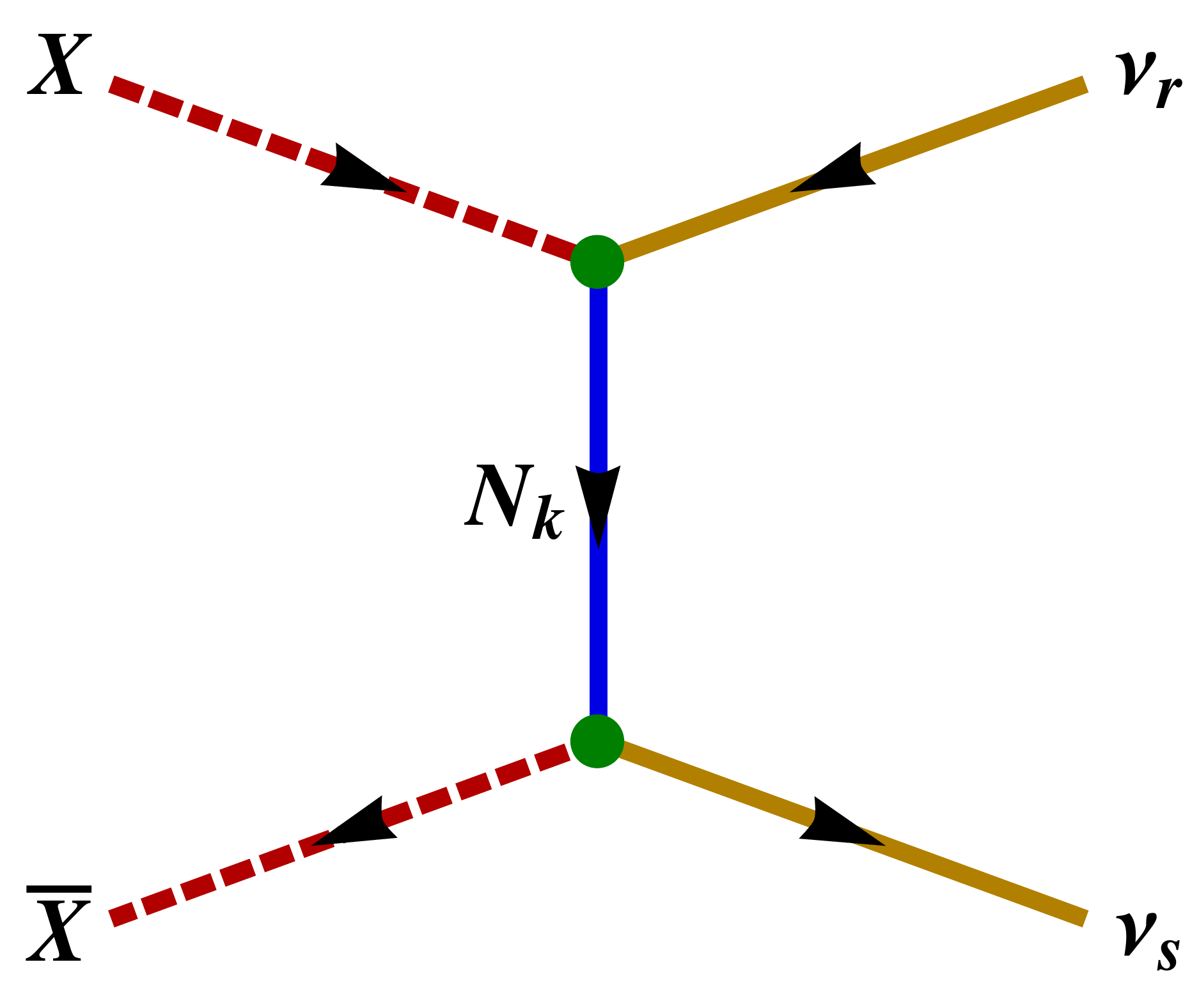}
\hspace{0.3cm}
\includegraphics[scale=0.18]{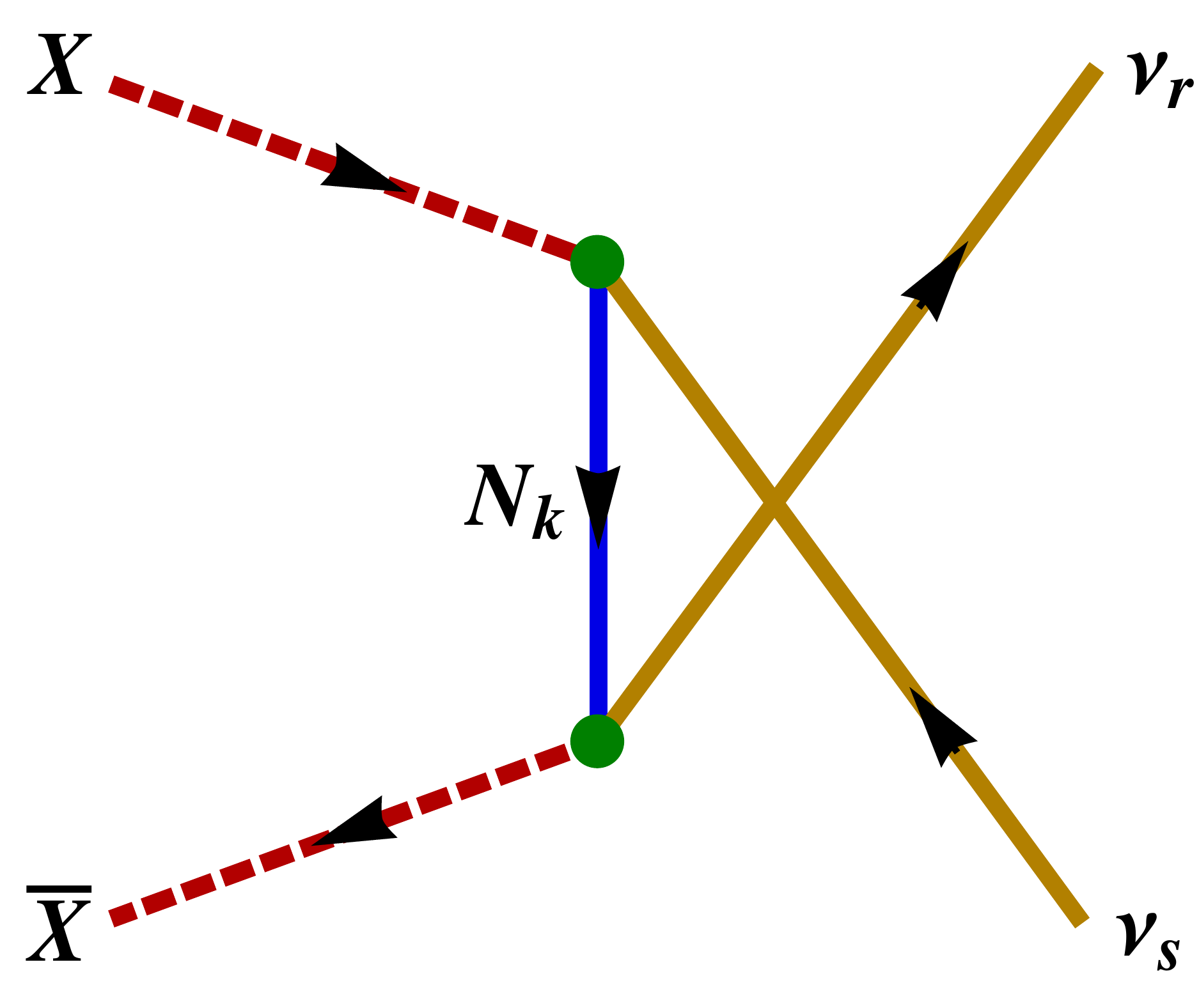} \\[0.1cm] $\hspace{0.19cm} (\text{a})$ $\hspace{7.36cm} (\text{b})$  
\vspace{-0.1cm}
\caption{(a) Feynman diagrams of the elastic scattering between $X$ and the SM neutrinos. 
(b) Feynman diagrams for the $2 \to 2$ annihilation process of a DM pair into a pair of the 
SM neutrino.}
\label{fig:elastic}
\end{center}
\vspace{-0.4cm}
\end{figure}

From Eq.\eqref{Yukawa}, the particle $X$ can interact with the active neutrinos via the Yukawa couplings ${\cal Y}_{rk}{}^{}$.\footnote{The particle $S$ can also have the ${}^{}3 \to 2{}^{}$ annihilation processes, but it can only interact with the SM sector through the Higgs portal. In this case, the reaction rate would be governed by the Higgs mass, and may be too small to keep kinetic equilibrium with the SM sector.} The Feynman diagrams of the elastic scattering between $X$ and the SM neutrinos are displayed in Fig.~\ref{fig:elastic}(a), and its reaction rate can be computed as
\begin{eqnarray}
\Gamma_\text{kin} 
{}^{}=\,  
n_{\nu} {}^{} 
\sum_{r,{}^{}s} 
\big\langle \sigma_{X\nu_r \to X\nu_s} \upsilon_\text{rel}\big\rangle 
~,
\end{eqnarray}
where the neutrino number density $n_\nu$ and the thermally averaged
effective scattering cross section are given by
\begin{eqnarray}\label{density}
n_\nu = 
\frac{3{}^{}{}^{}\hat{\zeta}(3)}{2{}^{}\pi^2}\,T^3 
~,\quad
\big \langle \sigma_{X\nu_r \to X\nu_s} \upsilon_\text{rel} \big\rangle 
{}^{} ={}^{} 
\frac{3{}^{}m_X^2s_\xi^4}{16{}^{}\pi}
\sum_{k,{}^{}l} \frac{\text{Re}\big({\cal Y}_{rk}^\ast {}^{} {\cal Y}_{rl} {}^{} {\cal Y}_{sk} {}^{} {\cal Y}_{sl}^\ast \big)}
{\big(M_k^2-m_X^2\big)\big(M_l^2-m_X^2\big)} 
\bigg(\frac{T}{m_X}\bigg)
~,
\end{eqnarray}
with $\hat{\zeta}(3) \simeq 1.202$ the Riemann zeta function of 3. 

\begin{figure}[t]
\vspace{-0.3cm}
\begin{center}
\includegraphics[scale=0.54]{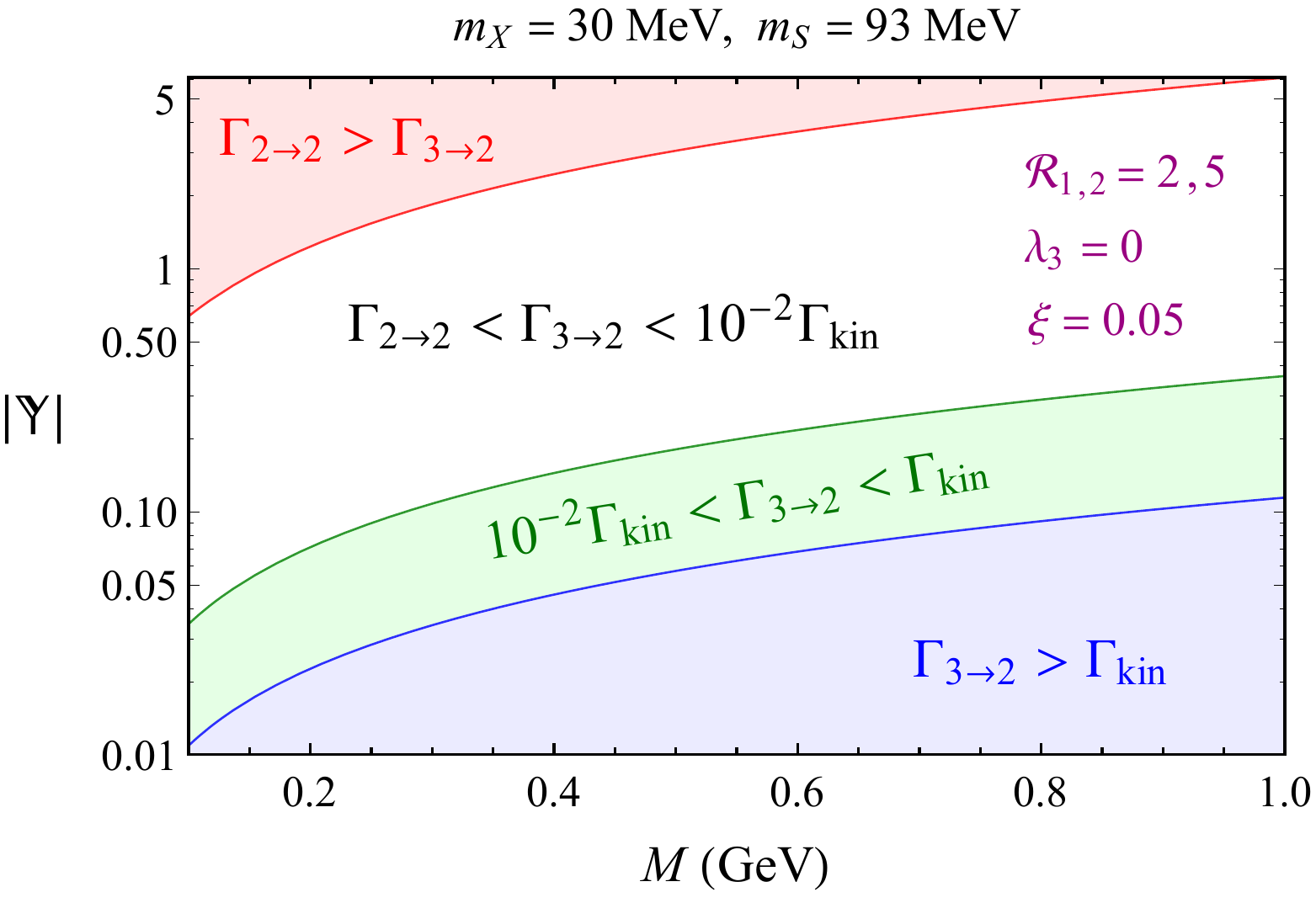}
\hspace{0.1cm}
\includegraphics[scale=0.54]{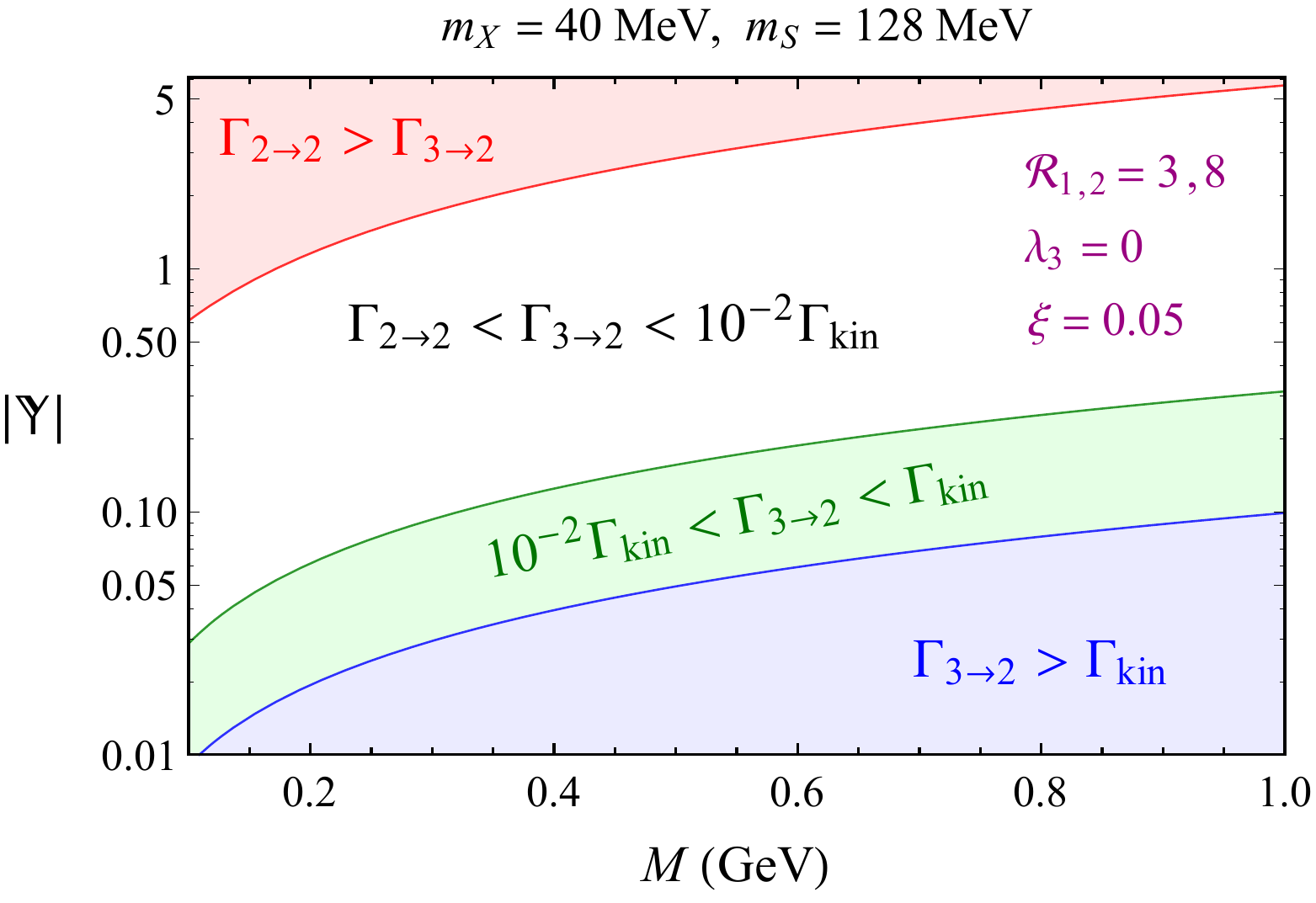}
\vspace{-0.3cm}
\caption{Magnitude of ${}^{}|\mathbb{Y}|$ versus $M$ for some choices of numerical sets. The white (red) region is the SIMP (WIMP) paradigm, the green region is the allowed parameter space of ${}^{}|\mathbb{Y}|$ by using the weaker SIMP condition, and the blue area is the failure of SIMP mechanism.}
\label{fig:SIMPcondition}
\end{center}
\vspace{-0.5cm} 
\end{figure}

By the crossing symmetry, the Feynman diagrams for the $2 \to 2$ annihilation process of a DM pair into a pair of the 
SM neutrino are shown in Fig.~\ref{fig:elastic}(b),\footnote{Here we have neglected the scattering processes $X \ell^{\pm} \to X \ell^{\pm}$ and the annihilation channels $X \bar{X} \to \ell^+\ell^-$ due to the mass suppression of the $Z$ boson and the Higgs boson.} and the reaction rate is calculated as
\begin{eqnarray}
\Gamma_{2 \to 2} \,=\, n_X 
\sum_{r,{}^{}s} \big\langle \sigma_{X\bar{X}\to {}^{}\nu_r\nu_s} \upsilon_\text{rel}\big\rangle
~, \quad
\end{eqnarray}
where the thermally averaged effective $2\to2$ annihilation cross
section is given by
\begin{eqnarray}\label{eq:ann}
\big\langle \sigma_{X\bar{X}\to {}^{}\nu_r\nu_s} \upsilon_\text{rel}\big\rangle 
\Eq
\frac{m_X^2s_\xi^4}{16{}^{}\pi}
\sum_{k,{}^{}l} \frac{\text{Re}\big({\cal Y}_{rk}^\ast {}^{} {\cal Y}_{rl} {}^{} {\cal Y}_{sk} {}^{} {\cal Y}_{sl}^\ast \big)}
{\big(M_k^2+m_X^2\big)\big(M_l^2+m_X^2\big)} 
\bigg(\frac{T}{m_X}\bigg)
~,\quad
\end{eqnarray}
with ${}^{}n_X{}$ given by Eq.\eqref{densityX}.  For simplification of numerical treatment, here we assume the masses of the 
vector-like fermions are degenerate\,($M_1 = M_2 = M_3 = M$). The reaction rates of the $2 \to 2$ annihilation process 
and the elastic scattering are then reduced to the form as
\begin{eqnarray}
\Gamma_{2 \to 2} 
\,=\, 
\frac{n_X {}^{} m_X^2 s_\xi^4}{16{}^{}\pi{}^{} x  \big(M^2+m_X^2\big)^{\hspace{-0.05cm}2}} {}^{}
\mathbb{Y}^4
~,\quad
\Gamma_\text{kin} 
\,=\, 
\frac{3{}^{} n_\nu {}^{} m_X^2 s_\xi^4}{16{}^{}\pi{}^{} x  \big(M^2-m_X^2\big)^{\hspace{-0.05cm}2}} {}^{}
\mathbb{Y}^4~,
\end{eqnarray}
where $\,\mathbb{Y} \equiv \scalebox{1.1}{\big[} {}^{}\sum_{r,{}^{}s,{}^{}k,{}^{}l} \text{Re}\big({\cal Y}_{rk}^\ast {}^{} {\cal Y}_{rl} {}^{} {\cal Y}_{sk} {}^{} {\cal Y}_{sl}^\ast \big) \scalebox{1.1}{\big]}^{\hspace{-0.05cm}1/4}$. Using the SIMP${}^{}{}^{}\,$condition  at $T_f$, we illustrate the plots of the magnitude of ${}^{}|\mathbb{Y}|$ as a function of ${}^{}M$ in Fig.~\ref{fig:SIMPcondition} with different numerical inputs based on Fig.~\ref{fig:Relic}. As indicated in the plots, 
the order of the Yukawa coupling is $|\mathbb{Y}| \sim {\cal O}(0.01-1)$ with $0.1 \,\text{GeV} \lesssim M \lesssim 1 \,\text{GeV}$.

\section{Conclusion}\label{sec:7}
In this paper, we have built a SIMP version of the scotogenic model,
where the SIMP DM has the responsibility to generate the neutrino masses
and its stability is guaranteed by the $\Zfive$ discrete symmetry. We
have considered the experimental and theoretical constraints
on the masses and the couplings in the model including the neutrino
masses and mixings, lepton flavor violating processes, anomalous magnetic moment, the invisible
decay modes of the $Z$ boson and the Higgs boson, the electroweak precision
data, perturbativity of the couplings and vacuum stability. 
In the models of SIMP DM, a large coupling is generally required in
order to reproduce the correct DM relic 
abundance measured by experiments through $\,3 {}^{}\to {}^{}2{}^{}{}^{}$ annihilating processes. 
This may give a tension with perturbativity and potential stability. 
By employing the resonant mechanism in our model, 
the correct relic abundance of DM has been reproduced, and the bounds on
the quartic couplings and the self-scattering cross section have been fulfilled at the same time. We found the parameter space of the new Yukawa interactions such that the SIMP condition is achieved.
Since our model faces to the stringent constraints from the Higgs invisible decay 
and the direct search of new charged scalars, it will be tested in near future.

\section*{Acknowledgments}
 T. T. acknowledges support from JSPS Fellowships for Research Abroad. 
The work of K.\,T is supported by JSPS Grant-in-Aid for Young Scientists (B) (Grants No.\,16K17697), by the MEXT Grant-in-Aid for Scientific Research on Innovation Areas (Grants No.\,16H00868), and by Kyoto University: Supporting Program for Interaction-based Initiative Team Studies (SPIRITS).

\vspace{0.5cm}

\appendix
\section{Gauged $\text{U}(1)_{\text{B}-\text{L}}$ extension of the $\nu$SIMP model}
It is believed that there is no global symmetry can exist in a theory of
quantum gravity~\cite{Ibanez:1991pr,Rai:1992xw}. Under this context, the
discrete symmetry we introduced in our $\nu$SIMP model may originate
from a gauge symmetry (gauge redundancy). At certain energy scale, this
gauge symmetry is broken down to the $\Zfive$ discrete symmetry by a
nonzero VEV of a scalar field. In the following, we demonstrate an extension of the $\nu$SIMP model to the gauged $\text{U}(1)_{\text{B}-\text{L}}\hspace{-0.05cm}$ version by adding one more SM singlet complex scalar $\zeta{}^{}$. The particle contents and the charge assignments are summarized in Tab.~\ref{tab:2}.

In this extended model, the Lagrangian associated with the $3 \to 2$ processes is given by
\begin{eqnarray}\label{zeta3to2}
{\cal L}_{\zeta} \,=\, 
\tfrac{1}{\sqrt{2}} \lambda_1  \zeta^\ast  \chi^\ast \hspace{-0.05cm} S^2 +
\tfrac{1}{\sqrt{2}} \lambda_2  {}^{}{}^{} \zeta^\ast  \chi^2 S \,+\, 
\tfrac{1}{6} \lambda_3 {}^{} \chi^3 S^\ast +
\text{H.c.} ~.
\end{eqnarray}
After spontaneous symmetry breaking, the complex scalar ${}^{}\zeta{}^{}$ can be expanded around its VEV 
as $\zeta = \tfrac{1}{\sqrt{2}}\big(\varsigma + \upsilon'\big)$, where $\upsilon' \equiv \sqrt{2} {}^{}{}^{} \langle \zeta \rangle$. 
The scalar interactions between $\chi$ and $S$ are then extracted as
\begin{eqnarray}\label{zeta3to2}
{\cal L}_{\zeta} \,\supset\, 
\tfrac{1}{2} \lambda_1  \upsilon' \chi^\ast \hspace{-0.05cm} S^2 +
\tfrac{1}{2} \lambda_2  {}^{}{}^{} \upsilon'  \chi^2 S \,+\, 
\tfrac{1}{6} \lambda_3 {}^{} \chi^3 S^\ast +
\text{H.c.} ~,
\end{eqnarray}
which corresponding to the first three terms in the last line of Eq.\eqref{V}, respectively, with $\mu_{1,2} = \lambda_{1,2} {}^{}{}^{} \upsilon'$. The Yukawa couplings contributed to the neutrino mass diagrams are the same with in Eq.\eqref{Yukawa}, and the lightest scalar particle involving in the diagrams can be a SIMP DM candidate. On the other hand, the SIMP condition can be achieved by $Z'$-portal instead of the Yukawa interactions due to the new gauge boson in this model. We leave the detailed study of the model to future work.

\begin{table}[htbp]
\begin{center}
\def\arraystretch{1.4}
\begin{tabular}[c]{|c||c|c||c|c|c|c|c|}
\hline               
~$\vphantom{|_|^|}$~ & ~$E$~ & ~$\Phi$~ & ~$N_{1,2, 3}$~ & ~$\eta$~ & ~$\chi$~ & ~$S$~ & ~$\zeta$~ \\[0.06cm]\hline\hline 
~SU(2)$\vphantom{|_|^|}$~       
& ~$\bm{2}$~ & ~$\bm{2}$~ & ~$\bm{1}$~ & ~$\bm{2}$~ & ~$\bm{1}$~ & ~$\bm{1}$~ & ~$\bm{1}$~ \\\hline
~U(1)$_{Y}\vphantom{|_|^|}$~      
& ~$-$1/2~ & ~1/2~ & ~0~ & ~1/2~ & ~0~ & ~0~ & ~0~ \\\hline
~U(1)$_{\rm{B}-L}\vphantom{|_|^|}$~      
& ~$-$1~ & ~0~ & ~$-$3/5~ & ~2/5~ & ~2/5~ & ~6/5~ & ~2~ \\\hline
\end{tabular}
\caption{Charge assignments of the particles in the gauged $\text{U}(1)_{\text{B}-\text{L}}$ extension of the $\nu$SIMP model.}
\label{tab:2}
\end{center}
\end{table}

\vspace{-0.8cm}

\section{Gauge interactions}\label{sec:appa}
\begin{figure}[htpb]
\begin{center}
\includegraphics[scale=0.2]{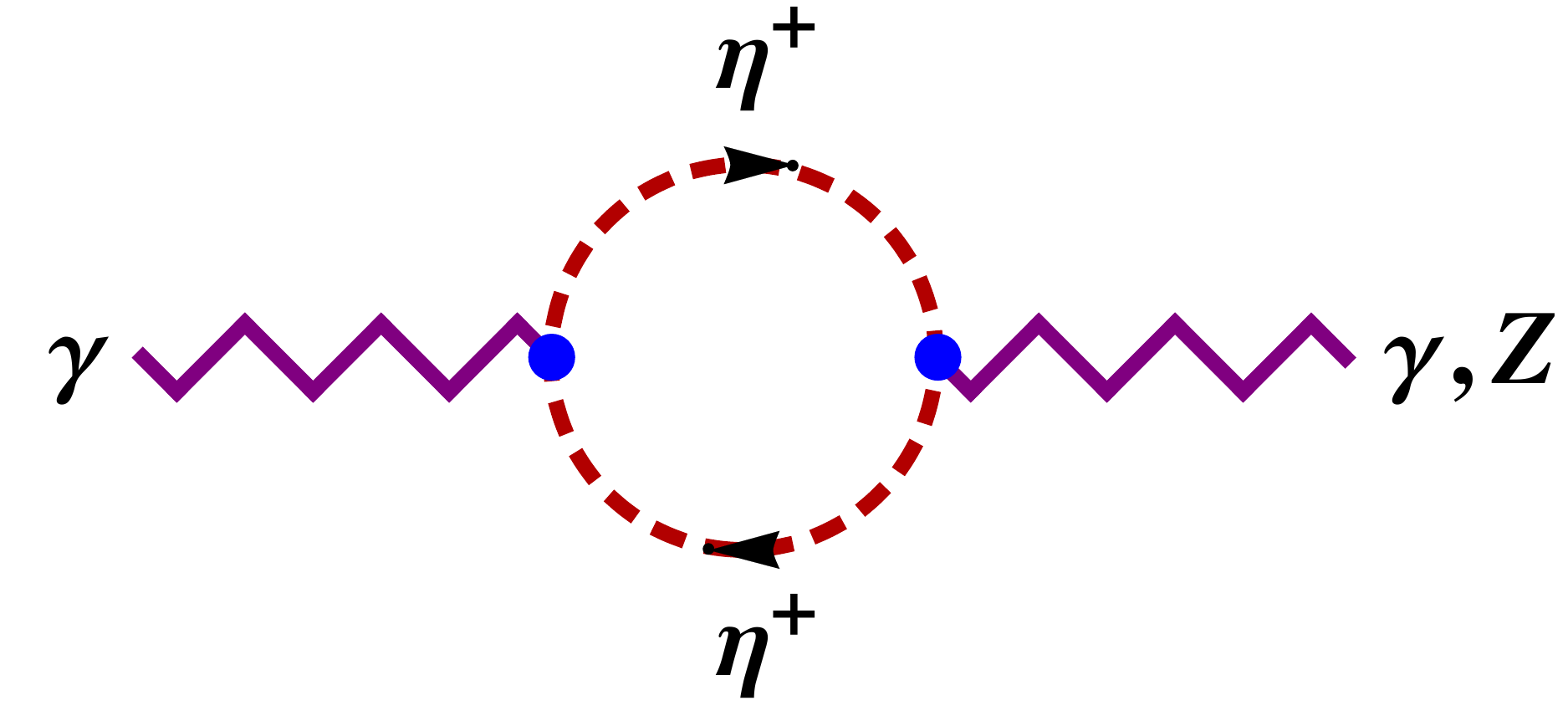}
\includegraphics[scale=0.2]{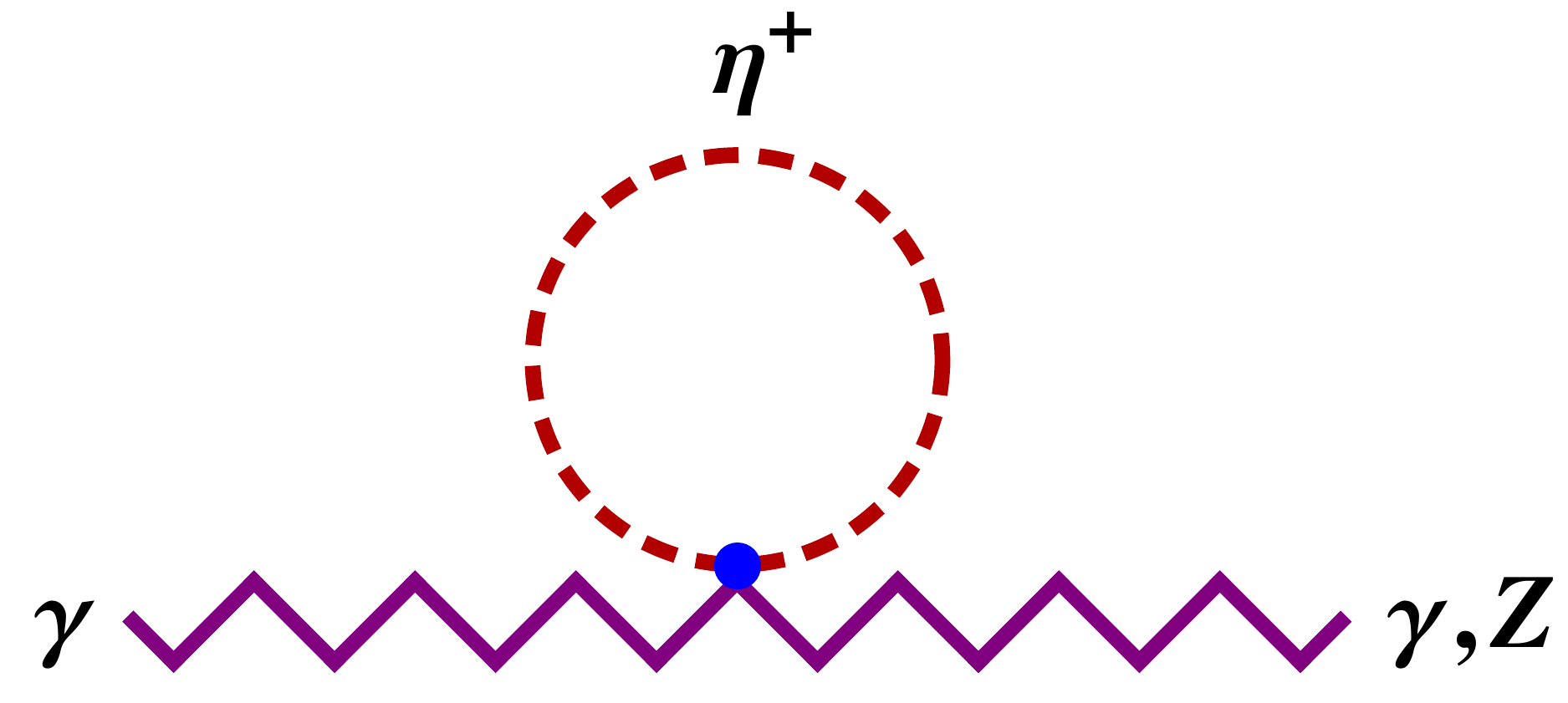}
\includegraphics[scale=0.2]{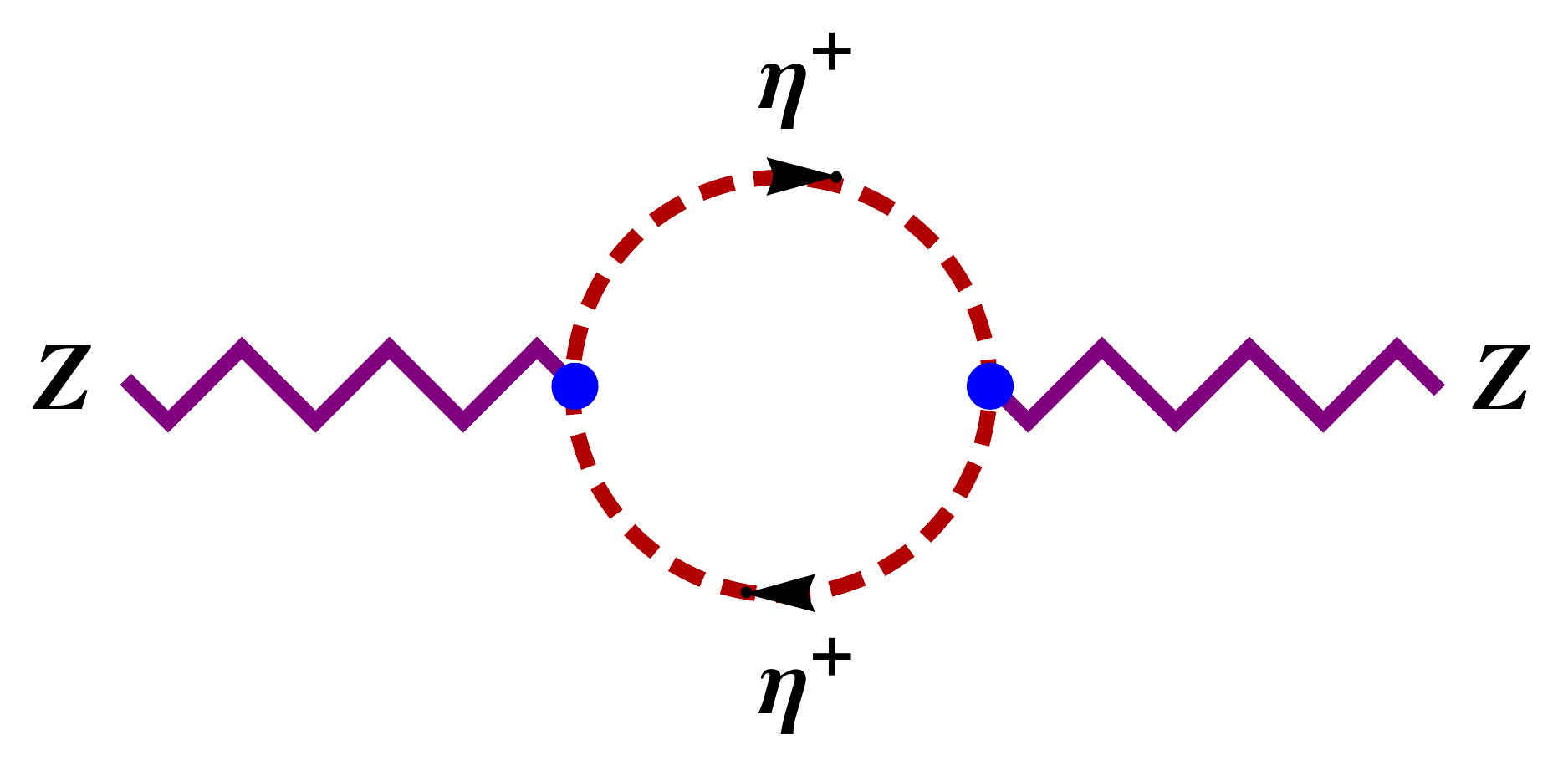}
\includegraphics[scale=0.2]{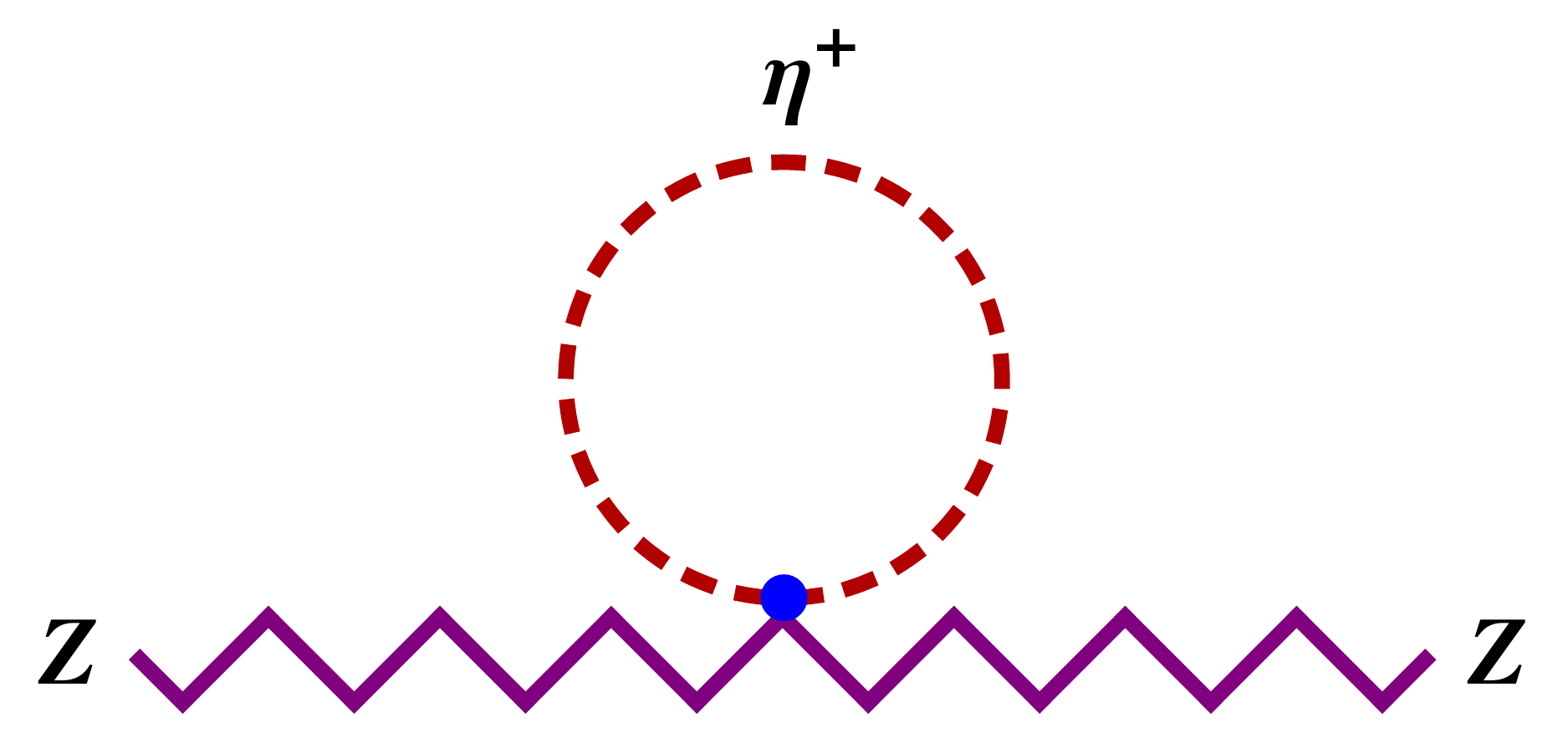} \\[0.3cm]
\hspace{0.15cm}
\includegraphics[scale=0.2]{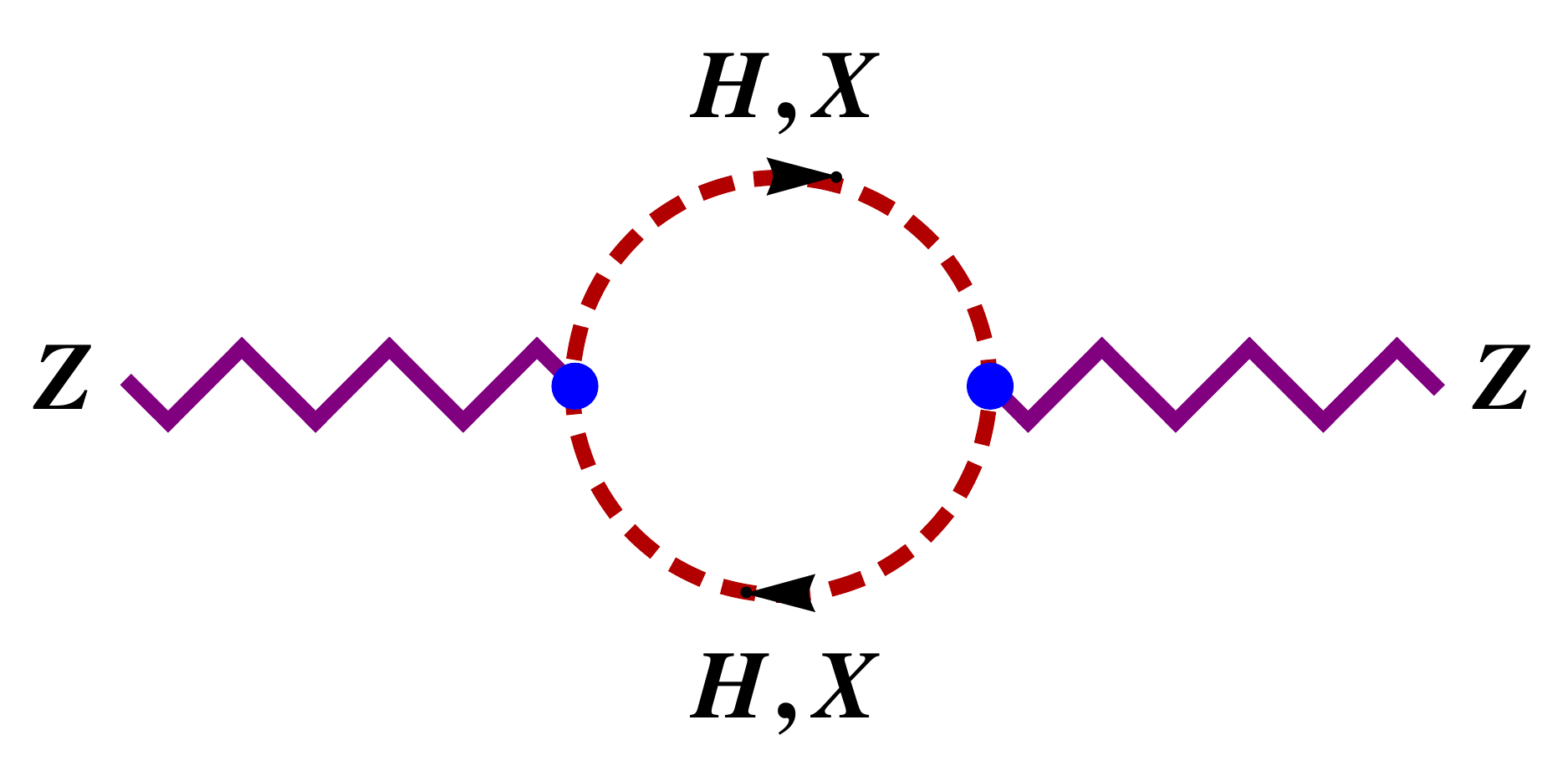}
\includegraphics[scale=0.2]{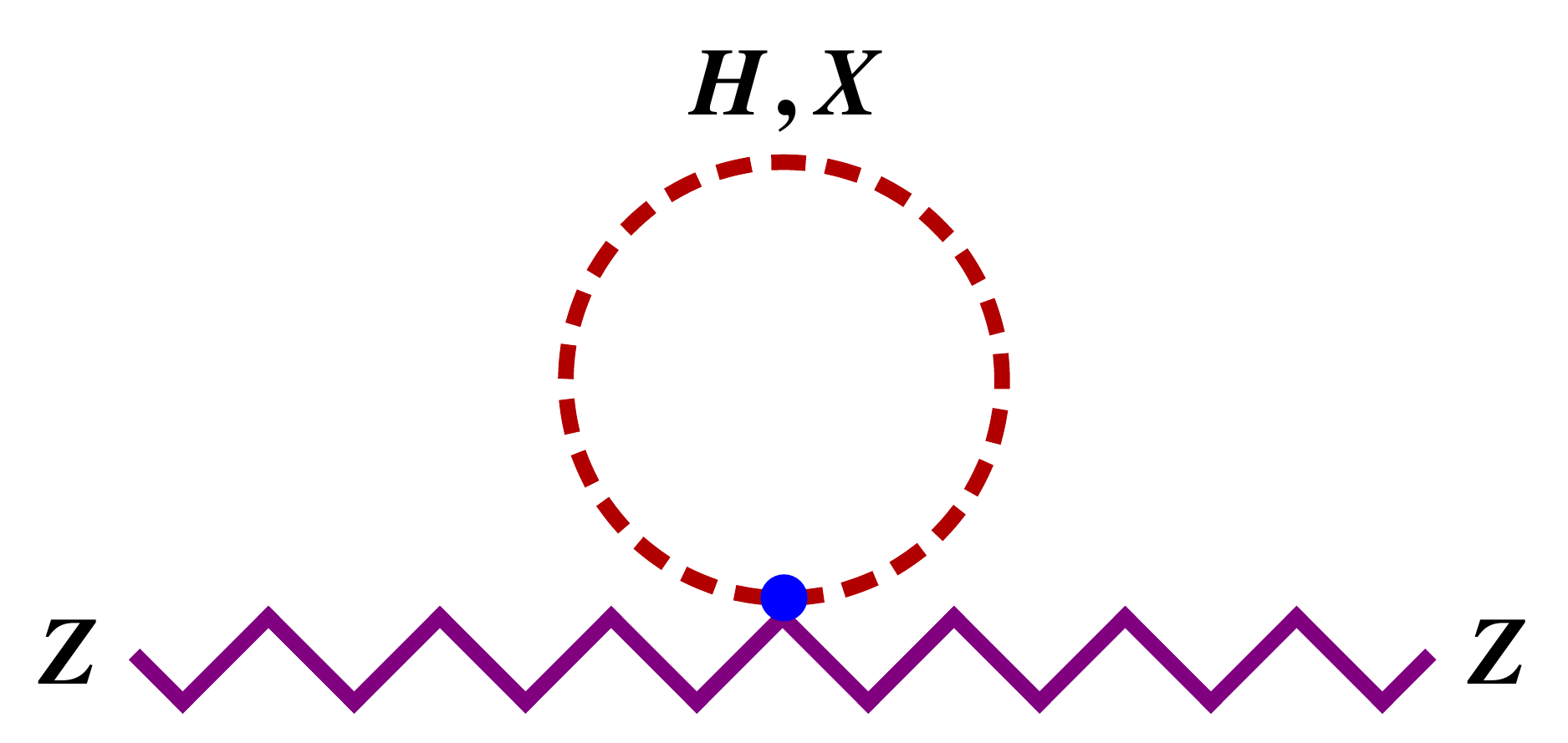}
\includegraphics[scale=0.2]{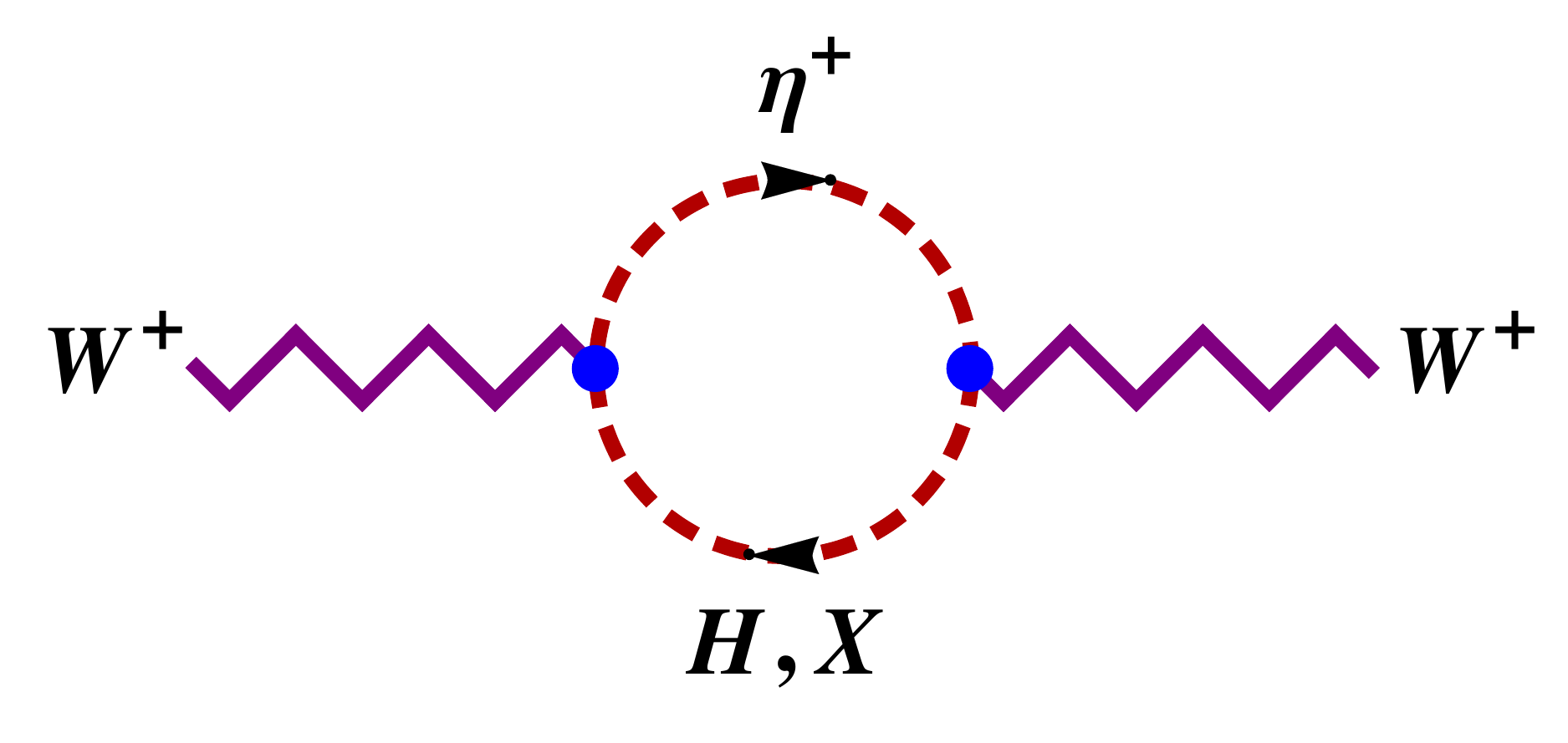}
\includegraphics[scale=0.2]{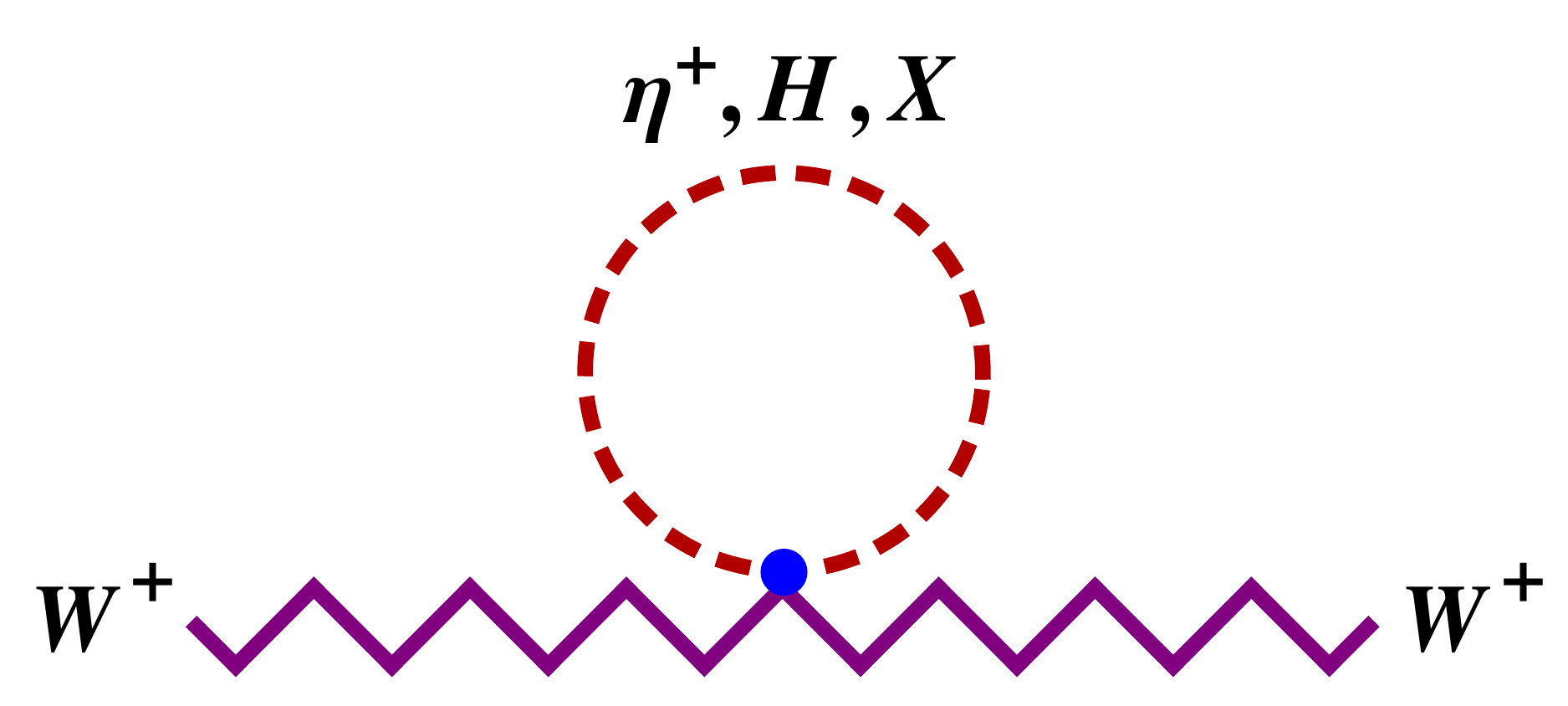}
\vspace{-0.2cm}
\caption{Feynman diagrams for the contributions of the new scalars to the oblique parameters $\Delta{\cal S},  \Delta{\cal T}$ and $\Delta {}^{} {\cal U}$.}
\label{fig:EW}
\end{center}
\end{figure}

The kinetic part of the Lagrangian in Eq.\eqref{LS} contains the interactions of the new scalars with the photon and
the weak bosons,
\begin{eqnarray}\label{gaugeint}
{\cal L}
&\,\supset\,&
i{}^{}\eta^+ \overset{\leftrightarrow}{\partial^\rho} \eta^-
\big(\hat{e} A_\rho + g_L Z_\rho \big) +
\frac{i{}^{}g_\text{w}}{2{}^{}c_\text{w}} 
\bigg[
c_\xi^2 {}^{} H^\ast \overset{\leftrightarrow}{\partial^\rho} H +
s_\xi^2 {}^{} X^\ast \overset{\leftrightarrow}{\partial^\rho} X +
c_\xi{}^{}s_\xi \Big( H^\ast \overset{\leftrightarrow}{\partial^\rho} X + X^\ast \overset{\leftrightarrow}{\partial^\rho} H  \Big)
\bigg]
Z_\rho 
\nonumber\\[0.1cm]
&&+\,
\frac{i{}^{}g_\text{w}}{\sqrt{2}}  
\bigg[
\Big(c_\xi{}^{} H \overset{\leftrightarrow}{\partial^\rho} \eta^-  + 
s_\xi {}^{}X \overset{\leftrightarrow}{\partial^\rho} \eta^- \Big) W^+_\rho
+
\Big(c_\xi {}^{}\eta^+ \overset{\leftrightarrow}{\partial^\rho} H^\ast  + 
s_\xi {}^{}\eta^+ \overset{\leftrightarrow}{\partial^\rho} X^\ast \Big) W^-_\rho
\bigg]
\nonumber\\[0.1cm]
&&+\,
\eta^+\eta^-\big(\hat{e}{}^{}A_\rho + g_L{}^{}Z_\rho\big)^{\hspace{-0.05cm}2} +
\frac{g_\text{w}^2}{4{}^{}c_\text{w}^2}
\Big[
c_\xi^2 |H|^2 + s_\xi^2 |X|^2 +
c_\xi s_\xi \big(H^\ast\hspace{-0.05cm} X+H X^\ast\big)
\Big] 
 Z^\rho Z_\rho 
\nonumber\\[0.1cm]
&&+\,
\frac{g_\text{w}^2}{2}
\scalebox{1.2}{\Big\{}
\eta^+\eta^-+
\Big[
c_\xi^2 |H|^2 + s_\xi^2 |X|^2 +
c_\xi s_\xi \big(H^\ast\hspace{-0.05cm} X+H X^\ast\big)
\Big] 
\scalebox{1.2}{\Big\}}
W^{+\rho} W^-_\rho  ~,
\end{eqnarray}
where
\begin{eqnarray}
{\cal W}\overset{\leftrightarrow}{\partial^\rho}{\cal X}
\,=\,
{\cal W}\partial^\rho {\cal X} - {\cal X} \partial^\rho {\cal W}
~, \quad
g_L = \frac{g_\text{w}}{2{}^{}c_\text{w}}\big(1-2{}^{}s_\text{w}^2\big) ~,\quad
s_\text{w} = \sqrt{1-c_\text{w}^2} ~.
\end{eqnarray}
With these gauge interactions, we draw the Feynman diagrams of the contributions to the SM gauge boson propagators in Fig.~\ref{fig:EW}.

\section{Benchmark points}
Assuming ${}^{}{\cal Y}^L\gg {\cal Y}^R{}^{}$ and the other parameter set
\begin{eqnarray}
m_H \,= \,m_{\eta^+} \,= \, 300 ~\text{GeV}  ~,\quad
\xi \,=\, 0.05 ~,\hspace{1.5cm}\nonumber\\
M_1\,=\,0.4~\text{GeV}   ~,\quad
M_2\,=\,0.6~\text{GeV}   ~,\quad
M_3\,=\,1~\text{GeV}  ~,\quad
\nonumber
\end{eqnarray}
two benchmark Yukawa couplings are given as 
\begin{eqnarray}
{\cal Y} 
&=& 
\left(\begin{array}{ccc} 
0.1  & 0 &  0   \\
0 & 0.3 &0   \\ 
0 & 0 & 0.5
\end{array}\right) ,\quad
{\cal Y}^L
= 
\left(\begin{array}{ccc} 
2.26  & 1.61 & 0.333  \\
1.61  & 1.82 & 0.989 \\
0.333 & 0.989 & 0.879
\end{array}\right)\hspace{-0.05cm}\times10^{-3} ,
\label{eq:bench1}
\end{eqnarray}
for ${}^{}m_X=30~\text{MeV}$, $m_S=93~\text{MeV}$, $\mu_{2}=150~\text{MeV}$, and 
\begin{eqnarray}
{\cal Y} 
&=& 
\left(\begin{array}{ccc} 
0.1  & 0 & 0 \\
0 & 0.2 & 0  \\ 
0 & 0 & 0.3
\end{array}\right) ,\quad
{\cal Y}^L
= 
\left(\begin{array}{ccc} 
1.07  & 1.14 & 0.261 \\
1.14  & 1.93 & 1.16  \\
0.261 & 1.16 & 1.15
\end{array}\right)\hspace{-0.05cm}\times10^{-3} ,
\label{eq:bench2}
\end{eqnarray}
for ${}^{}m_X=40~\text{MeV}$, $m_S=128~\text{MeV}$, $\mu_{2}=320~\text{MeV}$. 
One can check that Eq.~(\ref{eq:bench1}) and (\ref{eq:bench2}) satisfy the
SIMP condition $({}^{}{}^{}\Gamma_{3 \to 2}/\Gamma_{2 \to 2} \,\simeq 10^4$ and \,$\Gamma_\text{kin}/\Gamma_{3 \to 2} \,\simeq 10^3{}^{})$ as shown in the left and right panels of
Fig.~\ref{fig:SIMPcondition}, respectively.
These benchmark points give normal ordering neutrino mass eigenvalues
and mixing angles consistent with neutrino oscillation data. 
It is also possible to take benchmark parameter sets in the cases for
${}^{}{\cal Y}^L\ll{\cal 
Y}^R{}^{}$ and inverted hierarchy, though these are not
shown here.


\end{document}